\documentclass[aps,prl,twocolumn,titlepage,nofootinbib,longbibliography]{revtex4}
\usepackage{graphicx,physics}
\usepackage{color}
\usepackage{dcolumn}
\usepackage{latexsym}
\usepackage{CJKutf8}

\usepackage{hyperref,amssymb}
\usepackage{url}
\usepackage{color,soul}
\usepackage{graphicx}
\usepackage{multirow}
\usepackage{amsmath}
\newcommand{\beq}{\begin{eqnarray}}
\newcommand{\eeq}{\end{eqnarray}}
\usepackage{mathrsfs}
\usepackage{float,soul}
\usepackage[dvipsnames]{xcolor}
\usepackage{mathtools}
\usepackage{slashed}
\usepackage{physics}	
\usepackage{graphicx}   
\usepackage{epstopdf}
\usepackage{subfigure}  
\usepackage{hyperref}   
\usepackage{bbold}
\usepackage{wasysym}
\usepackage{feynmp}
\usepackage{hyperref}
\hypersetup{colorlinks,
}

\begin{document}

\title{\large Topological Defect Formation Beyond the Kibble-Zurek Mechanism\\ in Crossover Transitions with Approximate Symmetries}
\author{Peng Yang$^{1,2}$}
\author{Chuan-Yin Xia$^{3}$}
\author{Sebastian Grieninger$^{4}$}
\author{Hua-Bi Zeng$^3$}
\email{zenghuabi@hainanu.edu.cn}
\author{Matteo Baggioli$^{1,2}$}
\email{b.matteo@sjtu.edu.cn}
\address{$^1$Wilczek Quantum Center, School of Physics and Astronomy, Shanghai Jiao Tong University, Shanghai 200240, China}
\address{$^2$Shanghai Research Center for Quantum Sciences, Shanghai 201315,China}
\address{$^3$Center for Theoretical Physics , Hainan University, Haikou 570228, China}
\address{$^4$Center for Nuclear Theory, Department of Physics and Astronomy,
 Stony Brook University, Stony Brook, New York 11794-3800, USA}

\begin{abstract}
The formation of topological defects during continuous second-order phase transitions is well described by the Kibble-Zurek mechanism (KZM). However, when the spontaneously broken symmetry is only approximate, such transitions become smooth crossovers, and the applicability of KZM in these scenarios remains an open question.
In this work, we address this problem by analyzing both a weakly coupled Ginzburg-Landau model and a strongly coupled holographic setup, each featuring pseudo-spontaneous breaking of a global U(1) symmetry. In the slow quench regime, we observe a breakdown of the universal power-law scaling predicted by the Kibble-Zurek Mechanism. Specifically, the defect density acquires an exponential correction dependent on the quench rate, following a universal form dictated by the source of explicit symmetry breaking. Although these dynamics extend beyond the scope of the traditional KZM, we demonstrate that a generalized framework, that incorporates the effects of explicit symmetry breaking into the dynamical correlation length, remains valid and accurately captures the non-equilibrium defect formation across the entire range of quench rates.
\end{abstract}
\maketitle
\color{blue}\textit{Introduction} 
\color{black} -- Critical phenomena typically emerge at second-order phase transitions and critical points (CPs), where the correlation length diverges and dynamical processes slow down \cite{ma2018modern}. These phenomena are ubiquitous across modern physics, manifesting in systems ranging from superconductors to magnetic materials, but extending even beyond that \cite{RevModPhys.80.1275}. Importantly, the physics near critical points is largely independent of the macroscopic details, exhibiting universal behavior characterized by critical exponents and power-law scalings, which are determined solely by the nature of the critical point. For these reasons, critical points and dynamics around them are often described by universal frameworks \cite{RevModPhys.49.435}, which are based on the concept of spontaneous symmetry breaking (SSB) of continuous global symmetries \cite{10.21468/SciPostPhysLectNotes.11}, \textit{e.g.}, Ginzburg-Landau formalism \cite{HOHENBERG20151}.

It was first realized by Kibble in the context of cosmology \cite{Kibble_1976,KIBBLE1980183}, and later extended to condensed matter physics by Zurek \cite{Zurek1985,Zurek:1993ek}, that the formation of topological defects when a system is quenched across a second-order critical point also follows this same universality \cite{doi:10.1142/S0217751X1430018X}. More precisely, the number $\hat{N}$ of topological defects forming during this non-equilibrium process obeys a power-law scaling in terms of the quenching rate $\tau_Q$,
\begin{equation}
    \hat N \propto \tau_Q^{(d-D)\frac{\nu}{1+ z \nu}}, \label{eq1}
\end{equation}
where $D$ and $d$ are the number of spatial dimensions and the dimension of the defects formed, while $\nu$ and $z$ are the critical exponents that determine the universality class of the phase transition. 

Eq.~\eqref{eq1} is a direct prediction of the Kibble-Zurek mechanism (KZM) that has been experimentally verified in several systems, including colloidal packings \cite{doi:10.1073/pnas.1500763112}, superfluid helium \cite{Hendry1994,Ruutu1996,Bauerle1996}, liquid crystals \cite{doi:10.1126/science.251.4999.1336,doi:10.1126/science.263.5149.943},  and many more \cite{PhysRevLett.89.080603,PhysRevLett.96.180604,Ulm2013,Pyka2013,Weiler2008,Lamporesi2013,Chomaz2015,doi:10.1126/science.1258676,Ko2019,Keesling2019,Erne2018,Lee2024,tlv7-bvkh}. 

Even though the Kibble-Zurek mechanism has been extensively tested for continuous second order phase transitions, less is known about the dynamics and universality of defect formation across other types of critical (or pseudo-critical) points. More precisely, whether the KZ mechanism remains valid and predictive for more general critical phenomena remains an open question. In this direction, first order phase transitions, inhomogeneous phase transitions, quantum phase transitions with symmetry breaking bias, nonequilibrium phase transitions, quenches across tricritical points, transitions to discrete time crystalline phases and quantum multicritical points have been recently explored \cite{PhysRevLett.132.241601,pelissetto2025kibblezurekdynamicsfirstorderquantum,Zurek2000,PhysRevLett.127.115702,PhysRevLett.123.130603,Shu2025,PhysRevLett.125.095301,zhang2025observationnearcriticalkibblezurekscaling,wang2025drivencriticaldynamicstricitical,wang2025tricriticalkibblezurekscalingrydberg,jara2025kibblezurekmechanismdissipativediscrete,pelissetto2025outofequilibriumspinodallikescalingbehaviors,cpl_42_3_030203}.

In several physical systems, symmetries are only approximate, \textit{i.e.}, softly broken in an explicit way. In these scenarios, the spontaneous breaking of these symmetries leads to \textit{pseudo-critical points} or \textit{crossover phase transitions}, where the divergence of the correlation length is tamed and the would-be Goldstone modes become massive, yet remain light \cite{BURGESS2000193}. A prototypical example of this is chiral symmetry in quantum chromodynamics (QCD), which is spontaneously broken by the quark condensate, but also explicitly broken due to the finite quarks' mass. Condensed matter examples also exist and include pinned charge density waves \cite{RevModPhys.60.1129}, magnetic systems in external magnetic field, and more exotic situations (\textit{e.g.}, \cite{PhysRevLett.121.237201,Claude2025}).

To the best of our knowledge, the formation of topological defects across crossover phase transitions has yet to be explored in detail. Furthermore, the applicability of the Kibble-Zurek mechanism and the persistence of its universal characteristics (\textit{e.g.}, Eq.~\eqref{eq1}) in these scenarios remain unclear. This is the primary focus of our work.

\color{blue}\textit{Ginzburg-Landau model} 
\color{black} -- We first consider a two-dimensional system described by a complex scalar order parameter $\psi(\mathbf{x})$. The Ginzburg-Landau (GL) free energy can be written as
\begin{equation}
     \!\!F(\psi,h)\!=\!\int\!\! d^2x \left(\!\alpha |\psi|^2\!+\!\frac{\beta}{2}|\psi|^4\!+\!\gamma|\nabla\psi|^2\!-\!2\mathrm{Re}\!\left(h^*\psi\right)\!\right)\!, \label{free}
\end{equation}
where $h$ is an external field that explicitly breaks the global U(1) symmetry, $\psi \rightarrow e^{i \varphi}\psi$. For thermal phase transitions, $\alpha\propto(T-T_c)$ where $T$ is the temperature and $T_c$ is the critical temperature associated to the spontaneous breaking of the U(1) symmetry in the limit of $h\to0$.

\begin{figure}[ht]
\includegraphics[width=\linewidth]{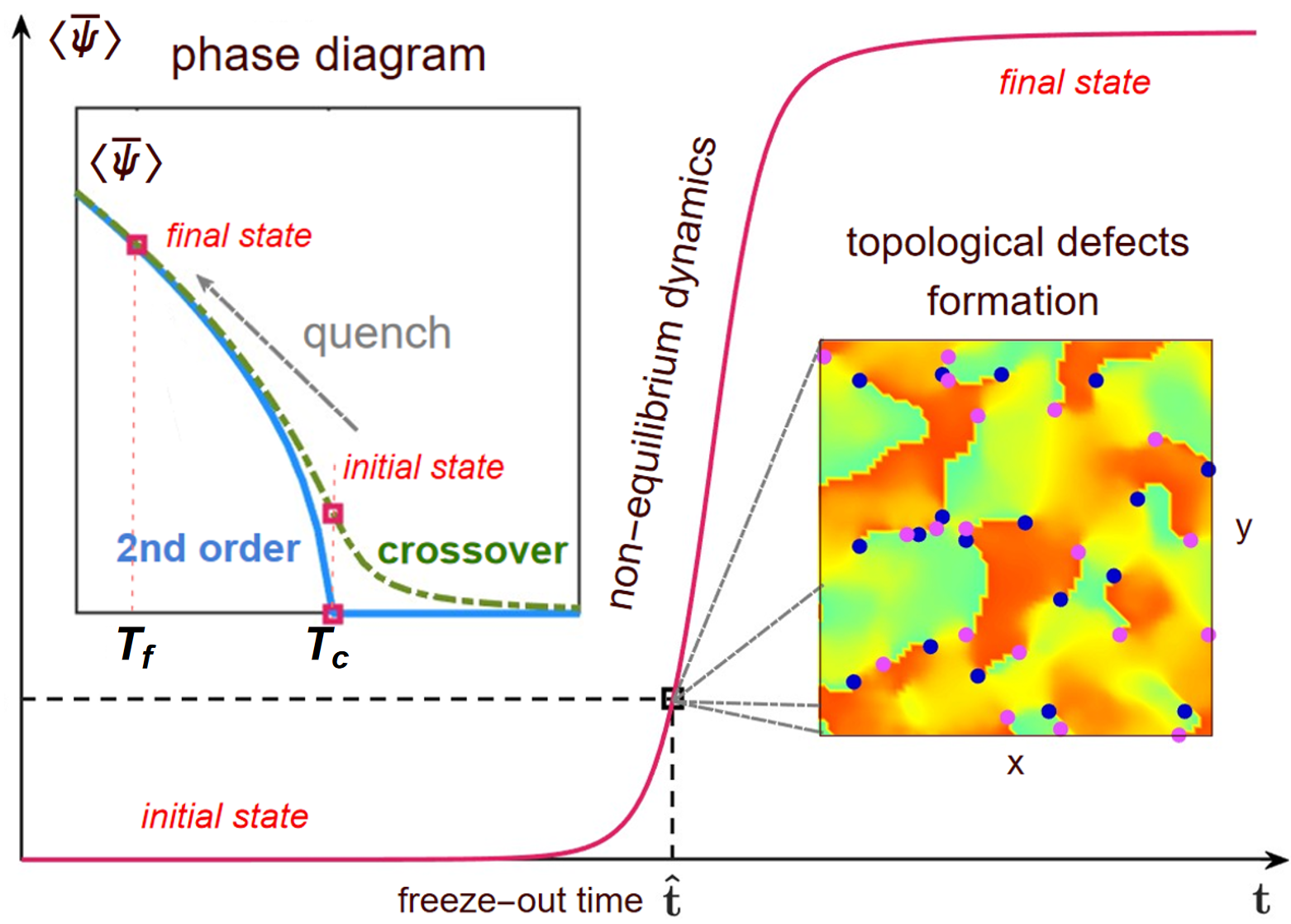}
\caption{\textbf{Quenching across a continuous crossover.} The spatially averaged expectation value of the order parameter $\langle \bar \psi\rangle$ as a function of time upon quenching the system from an initial state near the critical point $T=T_c$ to a final state at $T=T_f$. After a freeze-out time $\hat t$, the system starts to evolve following non-equilibrium dynamics in which the order parameter grows incoherently creating various spatial sub-regions with different phase $\theta=\arg (\psi)$. Topological defects form at the interfaces between these clusters whose size is determined by the correlation length $\xi$ (blue and purple symbols are respectively defects with positive and negative winding number). The top left inset shows the equilibrium value of $\langle \psi \rangle$ as a function of temperature across a continuous second order phase transition (blue line) as well as a continuous crossover (green line).} 
\label{fig1}
\end{figure}

In the $h=0$ limit, the homogeneous equilibrium solution can be found by minimizing the GL free energy, $\partial F/\partial\psi^*=0$, or equivalently $\alpha \psi+\beta|\psi|^2\psi=0$. Above the critical point, $\alpha>0$, the ground state equilibrium solution is $\langle \psi \rangle=0$ and the U(1) symmetry is not broken. At $T=T_c$, a second order phase transition characterized by the spontaneous breaking of the U(1) symmetry emerges. Below $\alpha=0$, in the broken (or superfluid) phase, we have a non-zero expectation value for the order parameter given by $\langle \psi\rangle=\sqrt{-\alpha/\beta}$ (see blue line in the phase diagram in Fig.~\ref{fig1}). When the parameter $h$ is non-zero, the minimization of the GL functional yields to $\alpha\psi+\beta|\psi|^2\psi=h$, and the sharp second order phase transition is now substituted by a continuous crossover (green line in the phase diagram in Fig.~\ref{fig1}). The crossover line can be obtained analytically and the expectation value of the order parameter $\langle \psi \rangle$ is now non zero for any value of $\alpha$. In the rest of the manuscript, we will always consider the \textit{pseudo-spontaneous} regime in which $h \ll \langle \psi \rangle$. The characteristics of the GL potential in Eq.~\eqref{free} for $h \neq 0$ are discussed in detail in the \textit{End Matter}, and they exhibit intriguing similarities with first-order phase transitions, which will be further elaborated in the \textit{Outlook} section.

In the $h=0$ limit, the critical point is described by two critical exponents $\nu$ and $z$, that in the mean-field approximation are fixed to $\nu=1/2$ and $z=2$. These fully characterize the critical dynamics and, in particular, the divergence of the relaxation time $\tau $ (\textit{critical slowing down}) and correlation length $\xi$, $\tau \propto |T-T_c|^{-\nu z}$ and $\xi \propto |T-T_c|^{-\nu}$. On the other hand, when $h \neq 0$, these divergent behaviors are regulated and both $\tau$ and $\xi$ are finite at any temperature (see \textit{End Matter}).

\color{blue}\textit{Non-equilibrium dynamics \& defect formation} 
\color{black} -- We consider a linear quench, $\alpha(t)=-t/t_Q$ from the critical point $\alpha=0$, to a final state corresponding to a temperature $T=T_f$ (see inset in Fig.~\ref{fig1}), where $\tau_Q$ defines the quench rate. The order parameter obeys the time-dependent Ginzburg-Landau equation
\begin{equation}\label{eq_TDGL}
    \partial_t \psi=-\Gamma\left(\alpha(t) \psi+\beta\psi|\psi|^2-\gamma\nabla^2\psi-h+\eta(t,\mathbf{x})\right),
\end{equation}
where $\Gamma$ is a phenomenological dissipation rate. Additionally, $\eta(t,\mathbf{x})$ is a delta-correlated random field (\textit{i.e.}, noise) with zero mean, satisfying
\begin{equation}
    \langle \eta(t,\mathbf{x})\eta(t',\mathbf{x'}) \rangle=\zeta\delta(t-t')\delta(\mathbf{x}-\mathbf{x'}),
\end{equation}
with $\zeta$ a small constant factor set to $\zeta=10^{-5}$. We numerically solve Eq.~\eqref{eq_TDGL} in a two-dimensional box. For more details, we refer to the Supplementary Material (SM).

Upon quenching the system, the spatially averaged order parameter $\langle \bar \psi\rangle$ evolves out-of-equilibrium, as shown in Fig.~\ref{fig1}. More precisely, the system does not respond up to a \textit{freeze-out} timescale $\hat{t}$, that we define as the time at which the condensate $\langle \bar \psi\rangle$ reaches $10\%$ of its final value $\langle \bar \psi(t \rightarrow \infty)\rangle$. We have verified that similar results are obtained by slightly changing this cutoff $\Lambda$ and also the initial configuration $\alpha(t=0)$ (see SM). 

During the out-of-equilibrium dynamics represented in Fig.~\ref{fig1}, the system displays the emergence of spatially coherent sub-domains where the order parameter takes the form $|\psi_i|e^{i\theta_i}$ with $i$ the index of the domain. A representation of this spatial structure is shown in the inset of Fig.~\ref{fig1}, where the colors indicate the value of the phase $\theta=\arg (\psi)$ in the two-dimensional box. Singularities in the phase $\theta$ emerge at the interfaces separating these sub-domains. These singularities correspond to topological defects with integer winding number $ \mathcal{W}=\frac{1}{2\pi} \oint_{\mathcal{L}} \theta d \mathbf{\ell}=\pm 1$. Defects with positive/negative charge are respectively vortices and anti-vortices, as indicated by blue and purple symbols in the inset of Fig.~\ref{fig1}. We can track numerically the number of defects formed during the out-of-equilibrium dynamics, $N,$ by inspecting the spatial structure of the order parameter at $t=\hat{t}$.

\begin{figure}[htb]
\centering
\includegraphics[width=\linewidth]{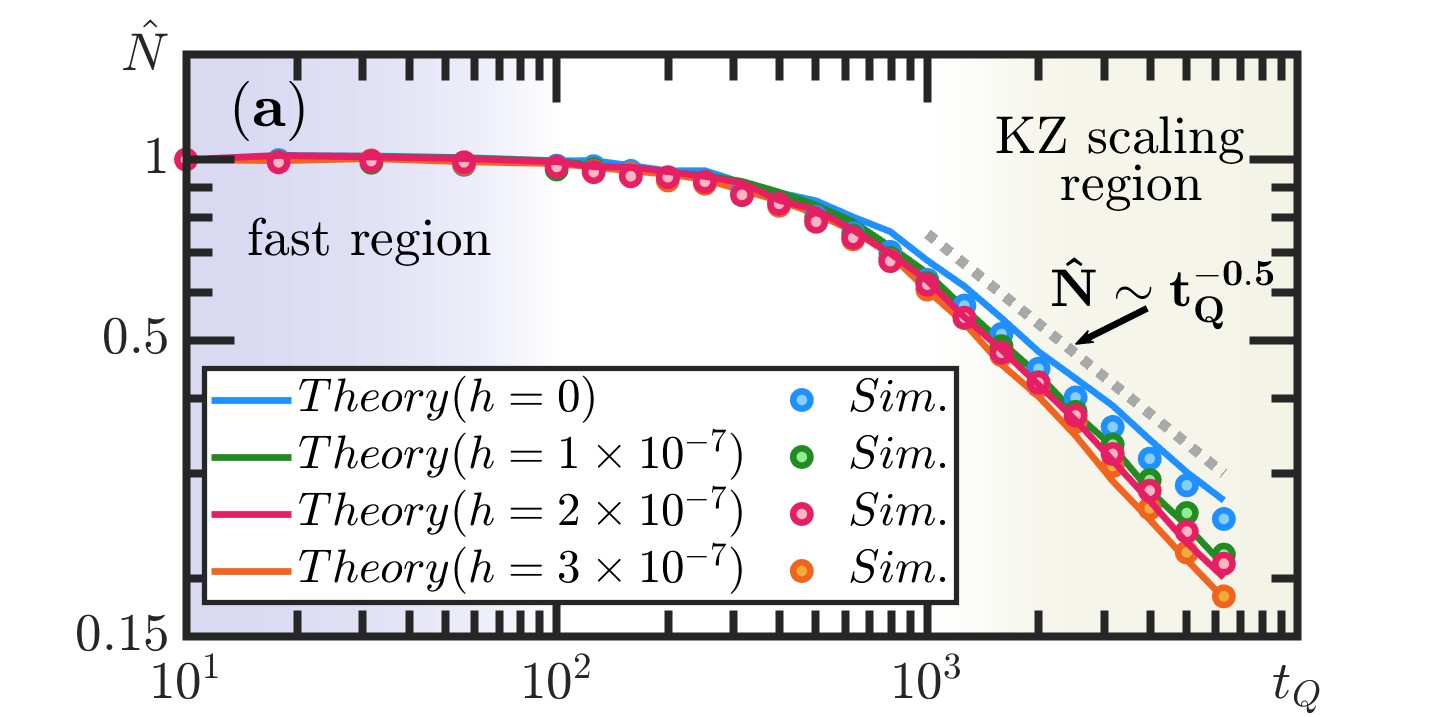}
\includegraphics[width=\linewidth]{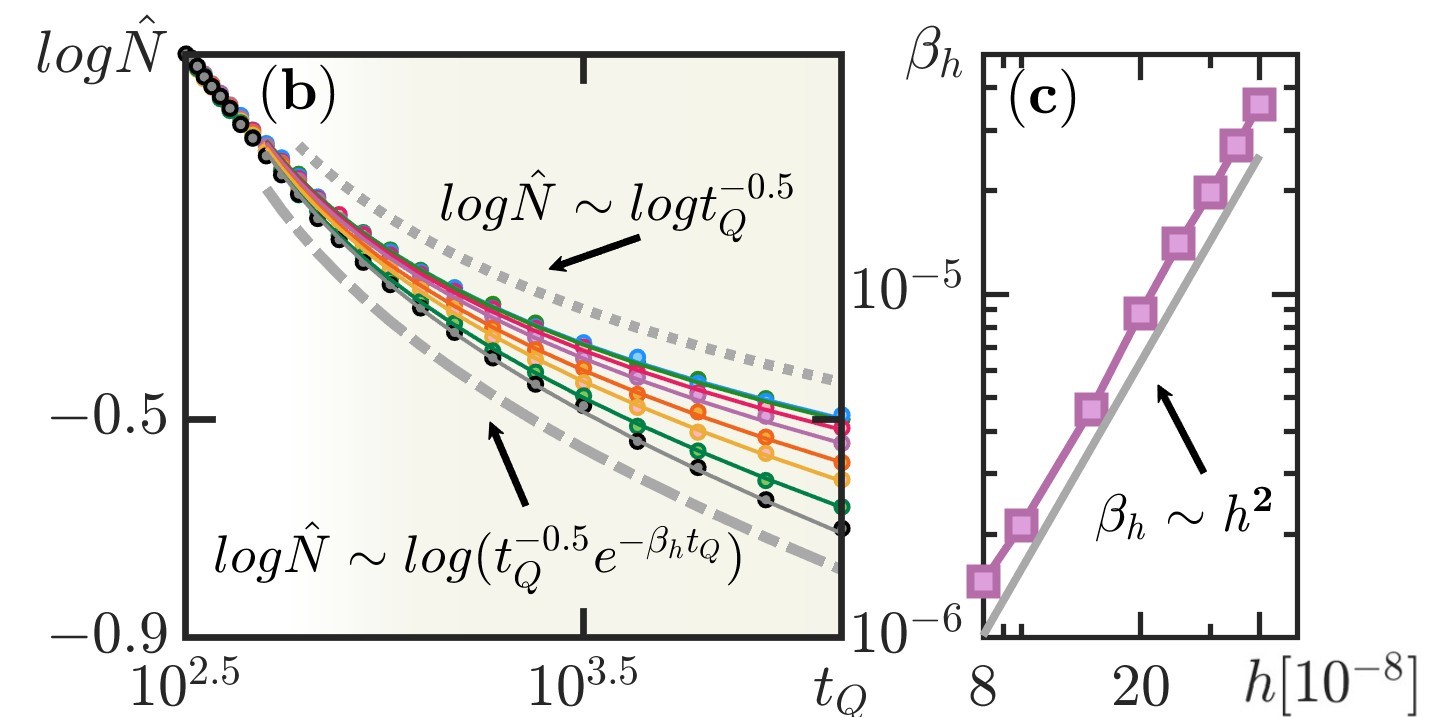}
\caption{\textbf{Ginzburg-Landau model.} \textbf{(a)} Normalized number of vortices $\hat{N}$ as a function of the quench rate $t_Q$. Background colors indicate respectively the fast quench region, the crossover regime and the KZ scaling region. Different colors correspond to different values of $h$. Symbols are the numerical data, solid lines the theoretical prediction introduced in the text, Eq. ~\eqref{scaling_law}. \textbf{(b)} Zoom in the KZ scaling region and fit of the numerical data using $\hat{N}\sim t_Q^{-0.5}e^{-\beta_h t_Q}$. \textbf{(c)} The behavior of $\beta_h$ as a function of the symmetry-breaking source $h$.}
\label{fig2}
\end{figure}

In Fig.\ref{fig2}(a), we show the normalized number of defects $\hat{N}\equiv N/N(t_Q\rightarrow 0)$ (colored symbols) as a function of the quench rate $t_Q$ for different values of the explicit breaking parameter $h$. At $h=0$ (blue symbols), the data can be separated into three regime: (i) A fast-quench regime ($t_Q <t_c \approx 10^2$) where $\hat{N}$ is independent of $t_Q$ (blue background color). The critical quench timescale \( t_c \) arises already in ideal second-order phase transitions, where it scales as $t_c \sim \epsilon_f^{-(z\nu + 1)}$,
with \( \epsilon_f \) quantifying the proximity of the final quench endpoint to the critical point and \( z \) and \( \nu \) denoting the critical exponents of the transition. More specifically, for thermal phase transitions one has \( \epsilon_f \sim (T_c - T_f) \), where \( T_f \) is the temperature reached at the end of the dynamical quench (see inset in Fig.~\ref{fig1}). Within the Ginzburg--Landau formalism, \( \epsilon_f \) is controlled by the value of the phenomenological parameter \( \alpha(t) \) reached at the end of the quench, which is set for simplicity to \( \alpha_f = -1 \) in all our simulations. The universal breakdown of KZ scaling in the fast-quench regime has been thoroughly analyzed and explained in Ref.~\cite{PhysRevLett.130.060402}, to which we refer for further details. (ii) An intermediate crossover region that interpolates between the fast and slow quench regimes ($10^2<t_Q<10^3$). (iii) A slow-quench regime, $t_Q>10^3$, where the number of defects follows the KZ scaling $\hat{N} \propto t_Q^{-0.5}$ (yellow background color).

By increasing the explicit breaking parameter $h$, several features emerge. First, the number of defects $\hat{N}$ decreases with $h$. As prove in detail in the following, this can be easily rationalized by the fact that the correlation length $\hat \xi$ increases with $h$. Intuitively, this can explained by the fact that explicit breaking tends to make the condensate uniform in space (see \textit{End Matter} for more details). In fact, in the limit of large $h$, the order parameter is totally homogeneous, leaving no space for the creation of topological defects, that become somehow ill-defined far away from the spontaneous symmetry breaking limit.

Most importantly, the behavior of $\hat{N}$ with $\tau_Q$ in the slow-quench regime is significantly affected by the presence of a symmetry-breaking source $h$. In particular, the original scaling law predicted by Kibble and Zurek no longer holds. A closer inspection, shown in Fig.~\ref{fig2}(b), reveals that the KZ scaling is modified as:
\begin{equation}
    \hat{N} \propto \tau_Q^{-1/2} e^{-\beta_h \tau_Q}, \label{eq2}
\end{equation}
with an exponential correction emerging due to the finite value of $h$. Equation~\eqref{eq2} provides an excellent fit to the numerical data.

In Fig.~\ref{fig2}(c), we show that the parameter $\beta_h$, which governs the exponential suppression of defect formation in the slow-quench regime, scales as $\sim h^2$ with the source strength introduced in the Ginzburg-Landau model. As expected, $\beta_h \to 0$ in the limit $h \to 0$, where the standard Kibble-Zurek mechanism is recovered.

\color{blue}\textit{Universality beyond weak coupling} 
\color{black} -- Our results so far are based on a phenomenological Ginzburg-Landau model which is valid only at weak coupling and in which dissipation and temperature are artificially introduced. In order to prove the universality of our findings beyond these limitations, we resort to a holographic model based on the AdS-CFT duality (see \cite{Liu_2020} for a relevant review in the context of nonequilibrium physics). In this framework, the KZM has been already discussed in several works and the KZ scaling has been numerically confirmed (\textit{e.g.}, { \cite{PhysRevX.5.021015,Sonner2015,xia2020winding,zeng2021topological}}).

Here, we extend those results by considering a holographic superfluid model in AdS$_4$ \cite{Herzog:2009xv,PhysRevLett.101.031601} in which the boundary global U(1) symmetry is also explicitly broken by a small source for a charged scalar operator. This model, already considered in \cite{Ammon:2021pyz,Donos:2021pkk}, realizes the pseudo-spontaneous breaking of the U(1) symmetry and exhibits a crossover phase transition as the one shown with green color in the inset of Fig.~\ref{fig1} (see SM for more details). More precisely, we consider the standard holographic superfluid model in four spacetime dimensions and within the probe limit \cite{Herzog:2009xv,PhysRevLett.101.031601} (see details in \textit{End Matter}). We notice that, in absence of any explicit breaking, the dynamics of this model are fully consistent with relativistic superfluid hydrodynamics \cite{Arean:2021tks} and the associated critical behavior is described by model F \cite{Flory:2022uzp,Donos:2022qao} in Hohenberg-Halperin classification \cite{RevModPhys.49.435}. This demonstrates the validity of this framework and its physical content. 

The holographic model is constructed in terms of a bulk gauge field that corresponds to a conserved current in the dual field theory description and a charged scalar field $\Psi(z,t,x,y)$ where $z$ is the holographic radial direction corresponding to the RG flow scale and $(t,x,y)$ the coordinates in the dual field theory. In the dual picture, this scalar field represents a charged scalar operator that, below a critical temperature $T_c$, spontaneously develops a finite expectation value, breaking spontaneously the global U(1) symmetry. 

\begin{figure}[htb]
\centering
\includegraphics[width=0.48\linewidth]{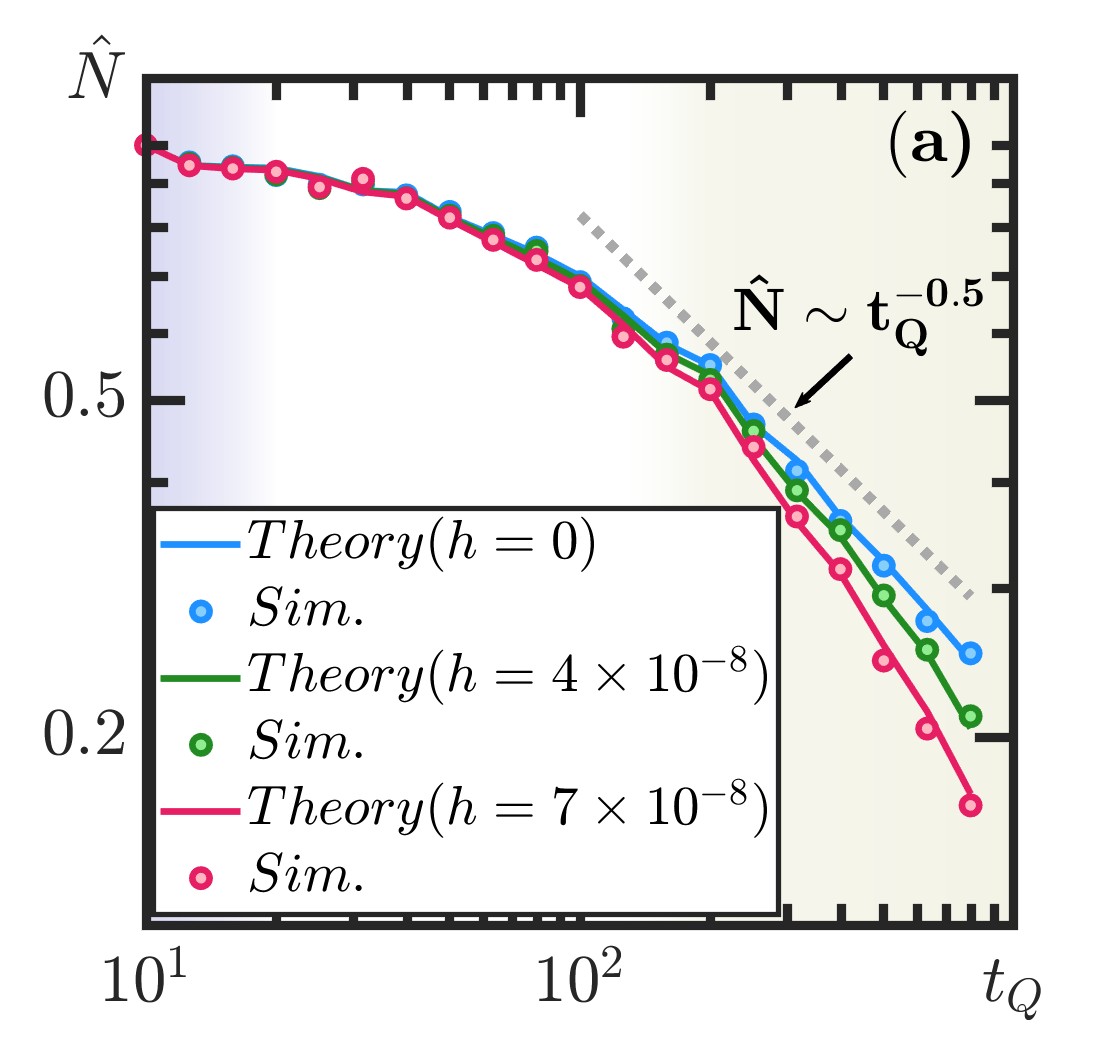}
\includegraphics[width=0.48\linewidth]{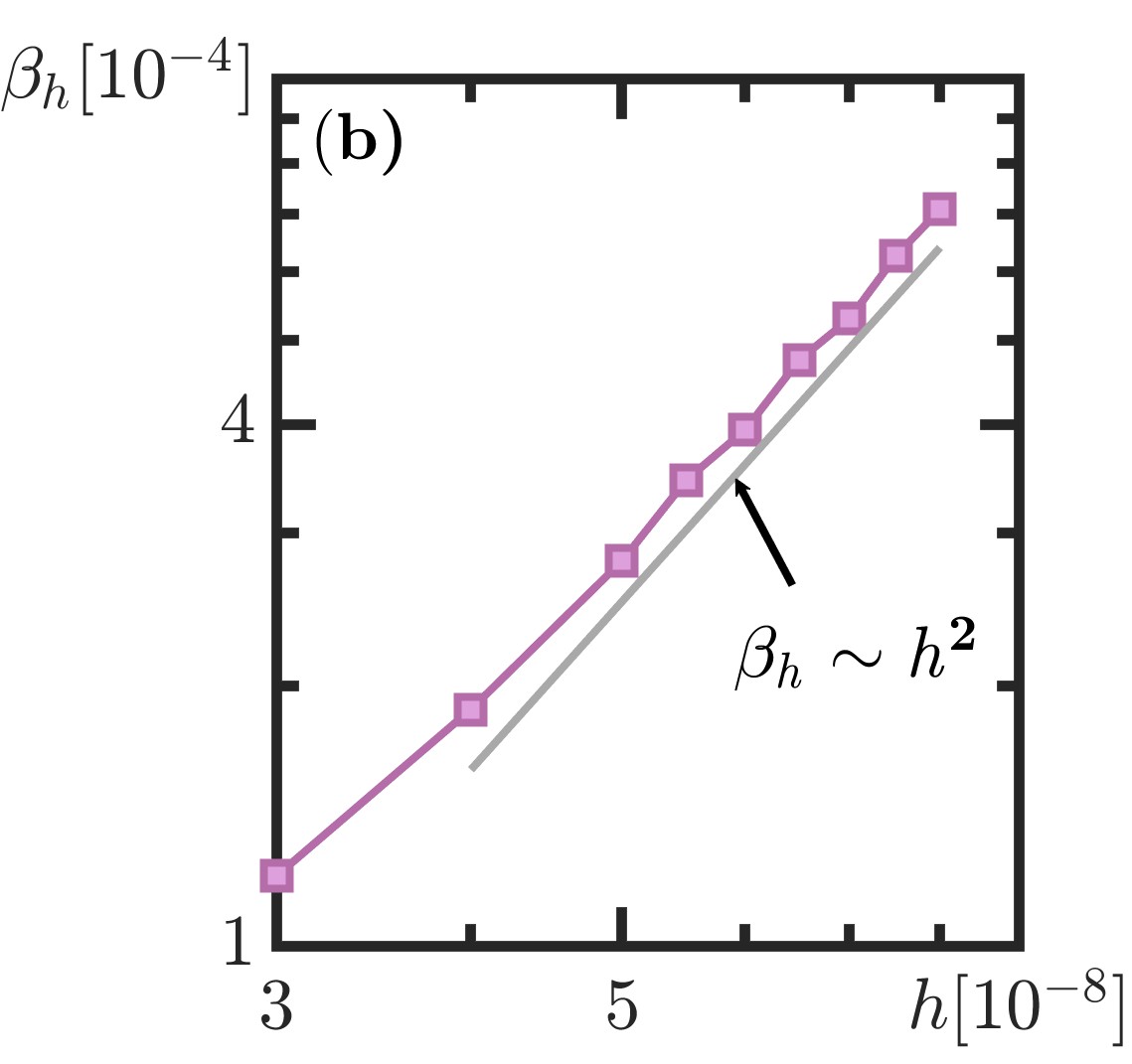}
\caption{\textbf{Holographic model.} \textbf{(a)} Normalized number of topological defects $\hat{N}$ formed during nonequilibrium quenches with inverse quench rate $t_Q$ in the strongly coupled holographic superfluid model. Symbols are the numerical data while solid lines are the theoretical predictions discussed in the text, Eq.~\eqref{scaling_law}. The dashed line guides the eyes towards the KZ scaling $\hat{N} \propto t_Q^{-0.5}$. \textbf{(b)} Verification of the scaling law $\beta_h\propto h^2$ by fitting the large $t_Q$ data using $\hat{N}\sim t_Q^{-0.5}e^{-\beta t_Q}$.}
\label{fig3}
\end{figure}

Following the same numerical protocol used in the GL model (see SM for details), we perform quenches with inverse rate $t_Q$ in the holographic setup and track the formation of topological defects throughout the nonequilibrium dynamics. In Fig.~\ref{fig3}(a), we present the normalized number of topological defects $\hat{N}$ as a function of the quench rate $t_Q$ for various values of $h$. The results exhibit the same qualitative behavior as observed in the GL model (cf. Fig.~\ref{fig2}), confirming the universality of our findings. Furthermore, as shown explicitly in Fig.~\ref{fig3}(b), the scaling $\beta_h \propto h^2$ holds in the holographic case as well, suggesting that this relation is universal and valid across both weakly and strongly coupled regimes.

\color{blue}\textit{Theoretical analysis \& generalized KZ mechanism} 
\color{black} -- To rationalize the observed deviations from the KZ scaling law and develop a generalized theoretical framework capable of capturing them, we analyze in detail the freeze-out time $\hat{t}$ and the corresponding spatial correlation length $\hat{\xi}$ in the GL model. We consider the density-density correlation function
\begin{align}\label{eq.G}
  \hat{G}(x-x',y-y') = \frac{1}{\rho^2} \langle \psi^*(x',y')\psi(x,y)\psi^*(x,y)\psi(x',y') \rangle,
\end{align}
evaluated at $t = \hat{t}$, where $\rho\equiv |\psi^2|$ denotes the condensate density. To extract the characteristic correlation length scale, we fit $\hat{G}(x-x', y-y')$ using a Gaussian function $e^{-(r - r')^2/\hat{\xi}^2}$, assuming rotational symmetry with $r \equiv (x,y)$. The parameter $\hat{\xi}$ defines the freeze-out correlation length and characterizes the average size of subdomains with coherent condensate (see inset in Fig.~\ref{fig1}). This ansatz yields a good fit to the numerical data; see \textit{End Matter} for details. We notice that in presence of a source $h$, $\hat{\xi}$ differs substantially from the equilibrium correlation length.

\begin{figure}[htb]
\centering
\includegraphics[width=\linewidth]{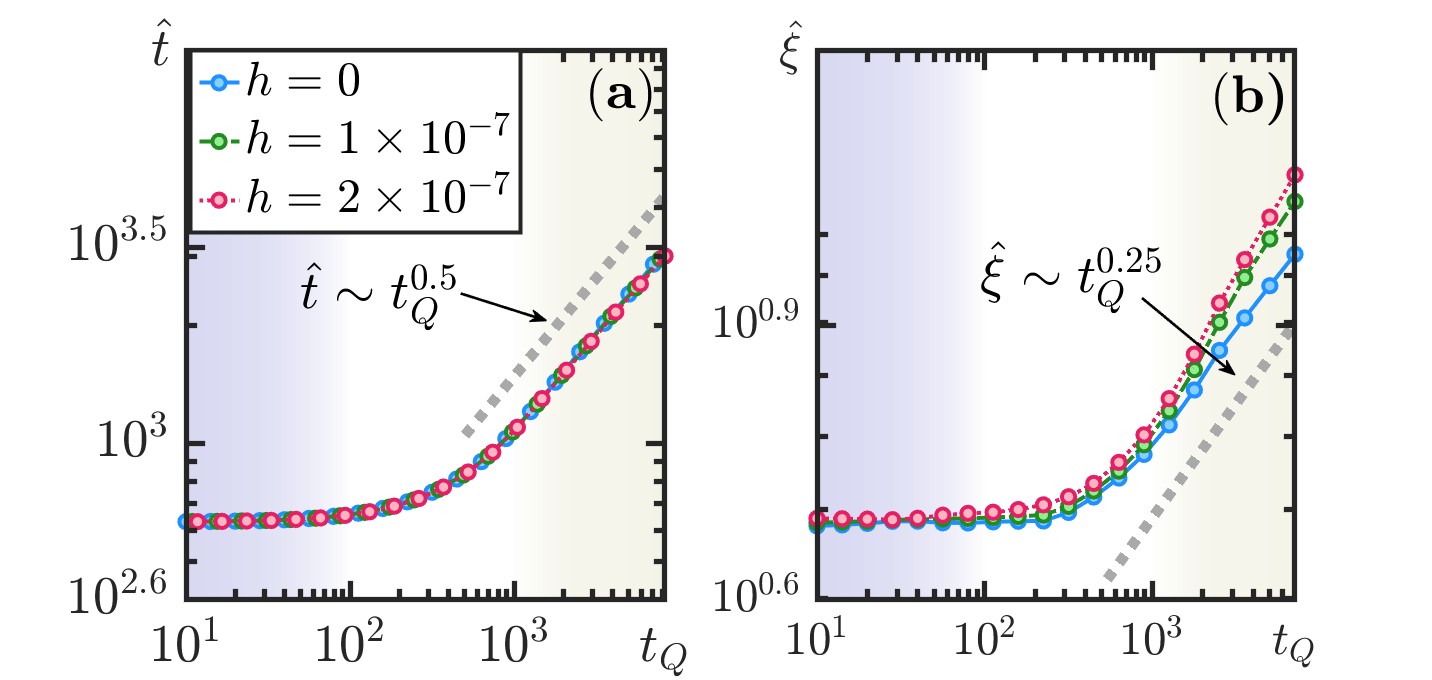}
\includegraphics[width=\linewidth]{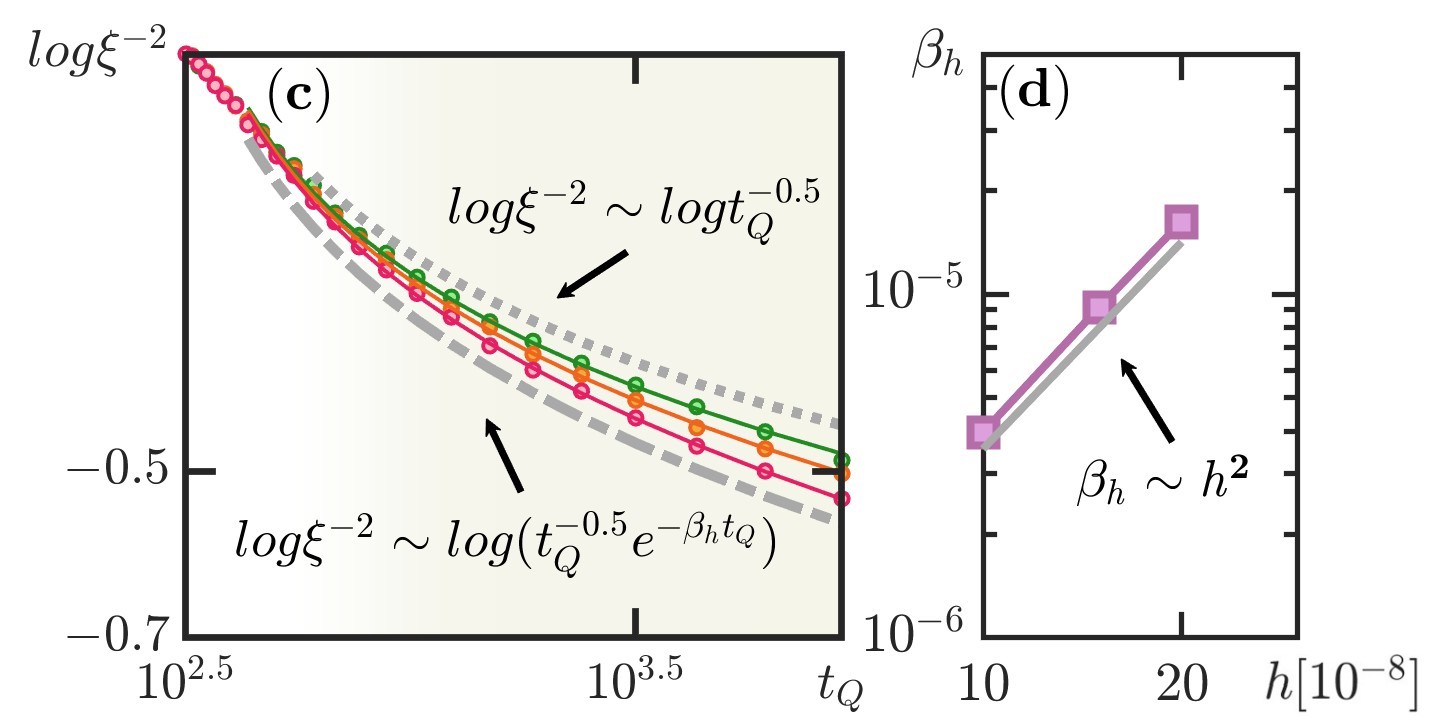}
\caption{\textbf{Ginzburg-Landau model.} \textbf{(a)} Freeze-out time $\hat{t}$ as a function of the quench rate $t_Q$ for different values of  $h$. The dashed line indicates the KZ scaling in the slow-quench regime. As explained in detail in the text, $\hat{t}$ is defined as the time at which the order parameter reaches $10\%$ of its final value. \textbf{(b)} The freeze-out correlation length $\hat{\xi}$ for the same data. \textbf{(c)} A zoom of $\hat{\xi}$ in the slow-quench regime displaying the deviations from the KZ predictions and fits to a generalized power-law scaling with exponential correction. $\beta_h=0$ corresponds to the KZ prediction. \textbf{(d)} Scaling of the parameter $\beta_h$ with $h$, confirming the origin of this trend.}
\label{fig4}
\end{figure}

In Fig.~\ref{fig4}(a), we show $\hat t$ as a function of $t_Q$ for three different values of the explicit breaking parameter $h$ within the GL model. Our results indicate that $h$ has, in first approximation, no effect on $\hat t$. In particular, for small enough $h$, all curves collapse on each other. Below a critical timescale $t_c \approx 10^2$, in the so-called \textit{fast quench} regime \cite{PhysRevLett.130.060402}, $\hat t$ is independent of $t_Q$. On the other hand, for slow quenches, $t_Q \gtrapprox 10^3$, the freeze-out time follows the Kibble-Zurek scaling $
\hat t \propto t_Q^{0.5}$ (dashed line), independently of the value of $h$. This scaling can be rationalized using the KZM \cite{doi:10.1142/S0217751X1430018X}.

More interesting results are found by examining the behavior of the freeze-out correlation length $\hat{\xi}$ when dialing $h$, see Fig.~\ref{fig4}(b). In particular, in the slow-quench regime this quantity exhibits strong deviations from the power-law $\sim t^{1/4}$ predicted by the KZ mechanism. Upon further inspection in Fig.~\ref{fig4}(c), these deviations can be parameterized by an exponential correction $\hat{\xi}\sim t^{1/4} \exp(\beta_h t_Q/2)$ where $\beta_h \propto h^2$ as demonstrated in panel (d) of the same figure.

This analysis reveals that the deviations from the power-law scaling predicted by the original KZ mechanism, observed in both the GL model (Fig.~\ref{fig2}) and the holographic model (Fig.~\ref{fig3}), originate from the behavior of the correlation length $\hat{\xi}$, and its deviation from the corresponding value at equilibrium usually employed in the traditional KZM. These findings suggest also the emergence of a generalized KZ mechanism capable of capturing this modified scaling beyond the standard power-law form.

In particular, we revisit the derivation of the defect number based on the Kibble-Zurek mechanism, which predicts
\begin{align}\label{scaling_law}
    \hat{N}= N_0 \frac{L^d}{\hat{\xi}^d},
\end{align}
where $L$ is the system size and, in our case, $d=2$. The prefactor $N_0$ is an undetermined constant, fixed by fitting the number of defects in the fast-quench regime, $N(t_Q \rightarrow 0)$, to Eq.~\eqref{scaling_law}. Crucially, and in contrast to the original KZM derivation, we do not assume any critical scaling behavior for the correlation length $\hat{\xi}$. Instead, we use the numerical values of $\hat{\xi}$ independently extracted from the correlation function at $t=\hat{t}$ and apply them in Eq.~\eqref{scaling_law}. As a result, Eq.~\eqref{scaling_law} becomes a parameter-free expression that yields a quantitative prediction for $\hat{N}$ as a function of $t_Q$.

The theoretical predictions are shown as solid lines in Fig.~\ref{fig2}(a) for the GL model and in Fig.~\ref{fig3}(a) for the holographic model. These results are in good agreement with the numerical data from our simulations, displayed using symbols of matching colors. This agreement confirms that the original Kibble-Zurek mechanism, encapsulated in Eq.~\eqref{scaling_law}, remains valid but must be refined by removing the assumption of critical scaling for the correlation length. While it would certainly be desirable to derive the exponential correction in $\hat{\xi}$ from first principles, our findings already indicate that a generalized KZ framework remains applicable even across crossover phase transitions. This result closely resembles recent experimental findings reported in~\cite{zhang2025observationnearcriticalkibblezurekscaling}, where deviations from the KZM were observed and effectively addressed by introducing the system size as a tuning parameter. Similar conclusions where reached using renormalization group arguments in \cite{PhysRevB.89.094108}.

\color{blue}\textit{Outlook} 
\color{black} -- In this work, we investigated the formation of topological defects during non-equilibrium quenches across continuous crossover phase transitions with an approximate U(1) symmetry. By combining a weakly coupled Ginzburg-Landau model with a strongly coupled holographic dual, we demonstrate the universality of the dynamics beyond the traditional Kibble-Zurek paradigm. 

We show that the explicit breaking of the U(1) symmetry induces an exponential suppression of the number of topological defects in the slow-quench regime, Eq.~\eqref{eq2}, leading to the breakdown of the original Kibble-Zurek (KZ) critical scaling law. Furthermore, we demonstrate that the strength of this exponential decay is universally governed by the magnitude square of the external source responsible for the symmetry breaking. These are the two main findings of our study.

We emphasize that, at this stage, our results are empirical numerical findings, and a rigorous theoretical foundation is still lacking. Interestingly, our results are consistent with biased quantum phase transitions~\cite{PhysRevLett.123.130603,PhysRevLett.127.115702}, as well as with quenches performed near, but not across, a critical point~\cite{10.1063/5.0191933}, where similar exponential corrections have been observed. As discussed in the \textit{End Matter}, the shape of the potential also bears intriguing similarities to first-order scenarios, where recent generalizations of the Kibble-Zurek (KZ) mechanism have been proposed \cite{PhysRevLett.132.241601}. Together, these parallels may serve as a valuable guide for developing a robust theoretical framework to explain our numerical observations.

Finally, we expect our findings to be universal with respect to the specific symmetry undergoing pseudo-spontaneous breaking. Consequently, we anticipate that similar behavior should manifest in a variety of systems, including the chiral phase transition in QCD, the critical dynamics of charge density waves in the presence of impurities, and magnetic systems subjected to external magnetic fields.

\color{blue}{\it Acknowledgments} \color{black} --  PY and MB acknowledge the support of the Shanghai Municipal Science and Technology Major Project (Grant No.2019SHZDZX01). SG was supported in part by a Feodor Lynen Research Fellowship of the Alexander von Humboldt Foundation. MB acknowledges the support of the sponsorship from the Yangyang Development Fund. PY is partially supported by the National Natural Science Foundation of China (Grant No. 12405021), the China Postdoctoral Science Foundation (Grant No. 2024T170545), and the Shanghai Post-doctoral Excellence Program (Grant No. 2024380).  HBZ acknowledges the support from the National Natural Science Foundation of China (under Grant No. 12275233 and 12547103). The computations in this paper were run on the Siyuan-1 cluster supported by the Center for High Performance Computing at Shanghai Jiao Tong University.

\section{End Matter}\label{end}
{\it Appendix A: Ginzburg-Landau potential and stationary points -} Assuming homogeneous configurations, the GL functional in Eq.~\eqref{free} in the main text can be conveniently rewritten in terms of the real and imaginary parts of the order parameter, $\psi=\psi_R+i \psi_I$, as
\begin{equation}
 F[\psi_R,\psi_I]= \alpha \left(\psi_R^2+\psi_I^2\right)+\frac{\beta}{2}\left(\psi_R^2+\psi_I^2\right)^2-2 h \psi_R, \label{pote}
\end{equation}
where the last term clearly breaks explicitly the U(1) symmetry, that is a global rotation in the plane $\left(\psi_R,\psi_I\right)$.

\begin{figure}[h]
\centering
\includegraphics[width=0.49\linewidth]{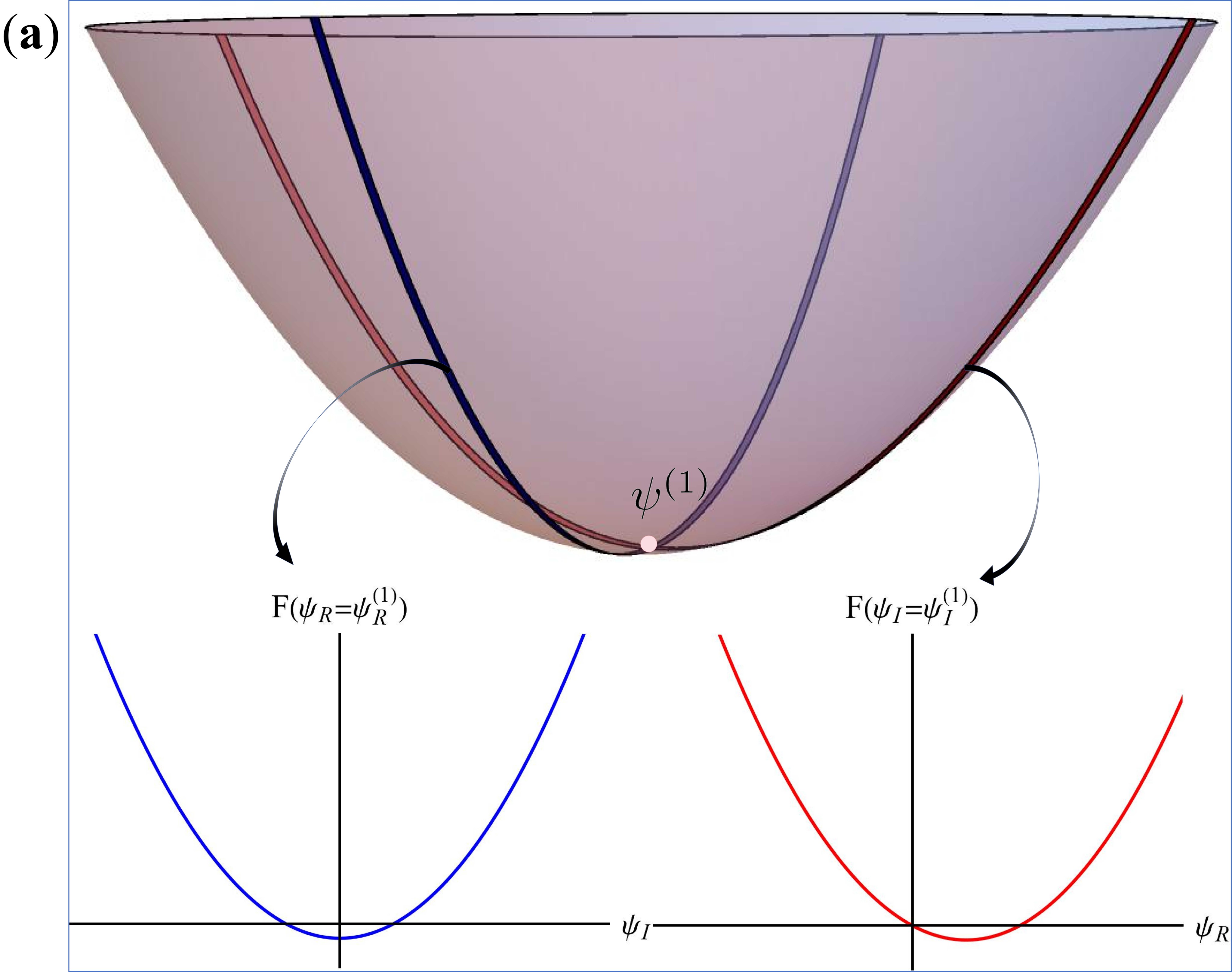}
\includegraphics[width=0.49\linewidth]{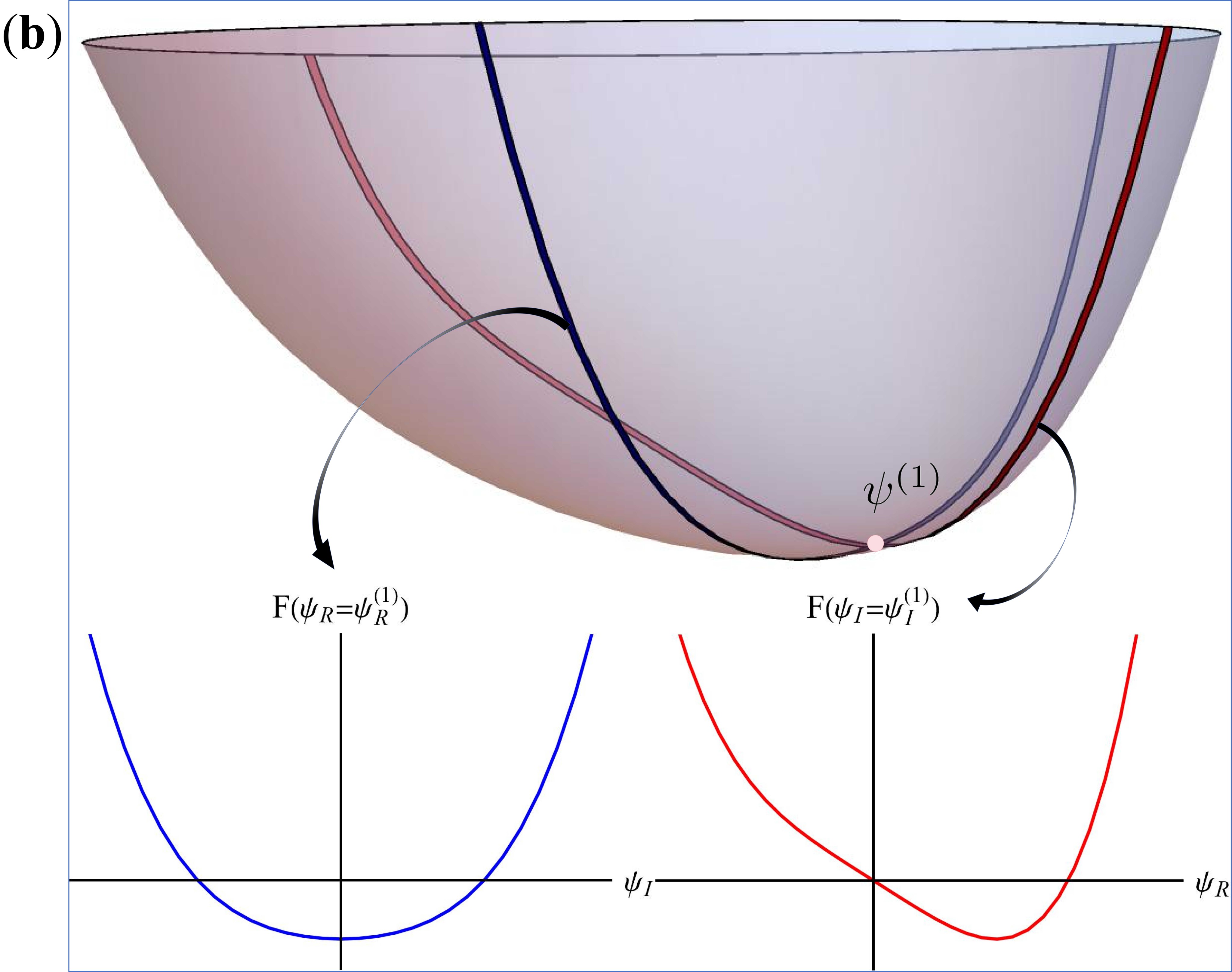}

\vspace{0.2cm}

\includegraphics[width=0.49\linewidth]{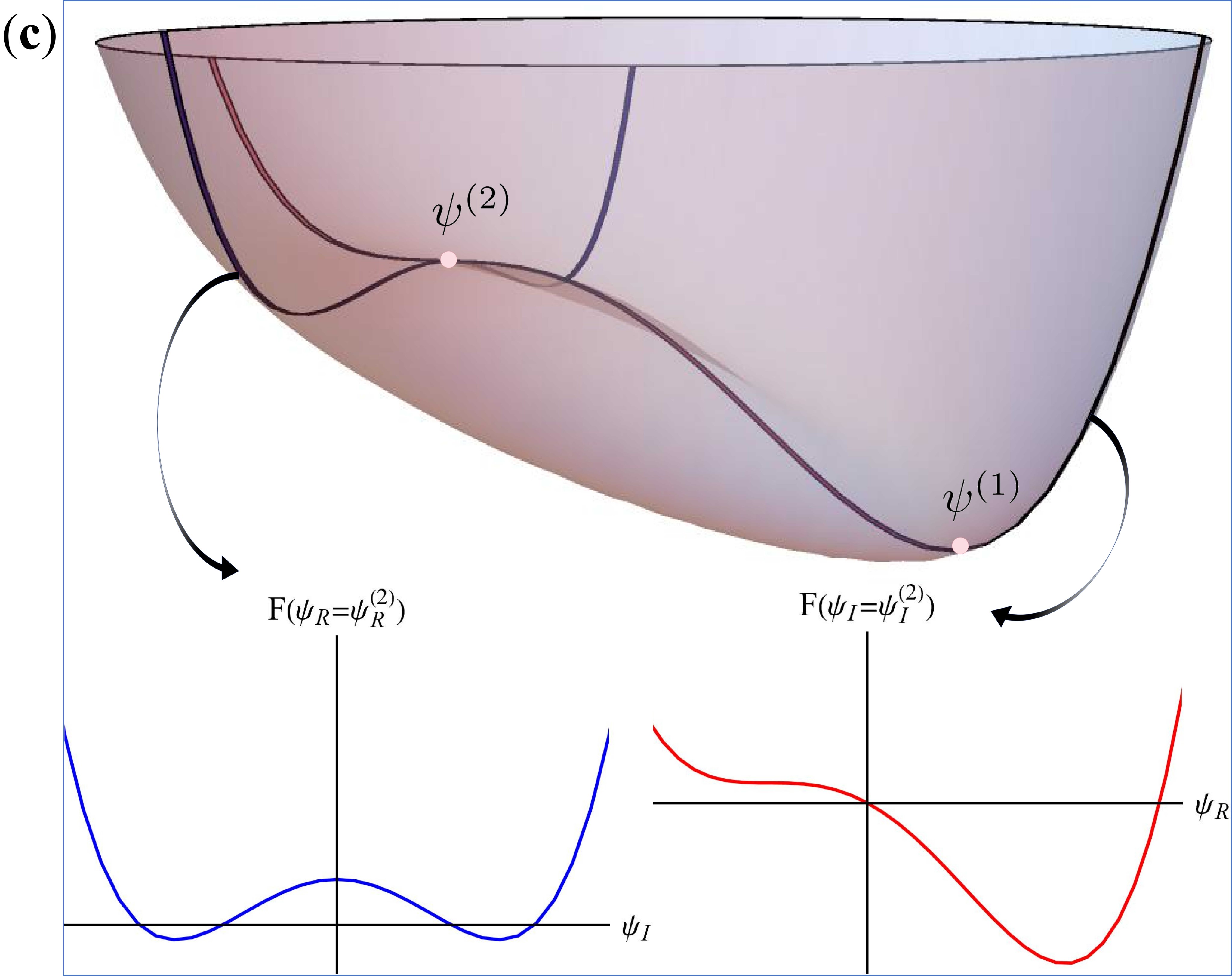}
\includegraphics[width=0.49\linewidth]{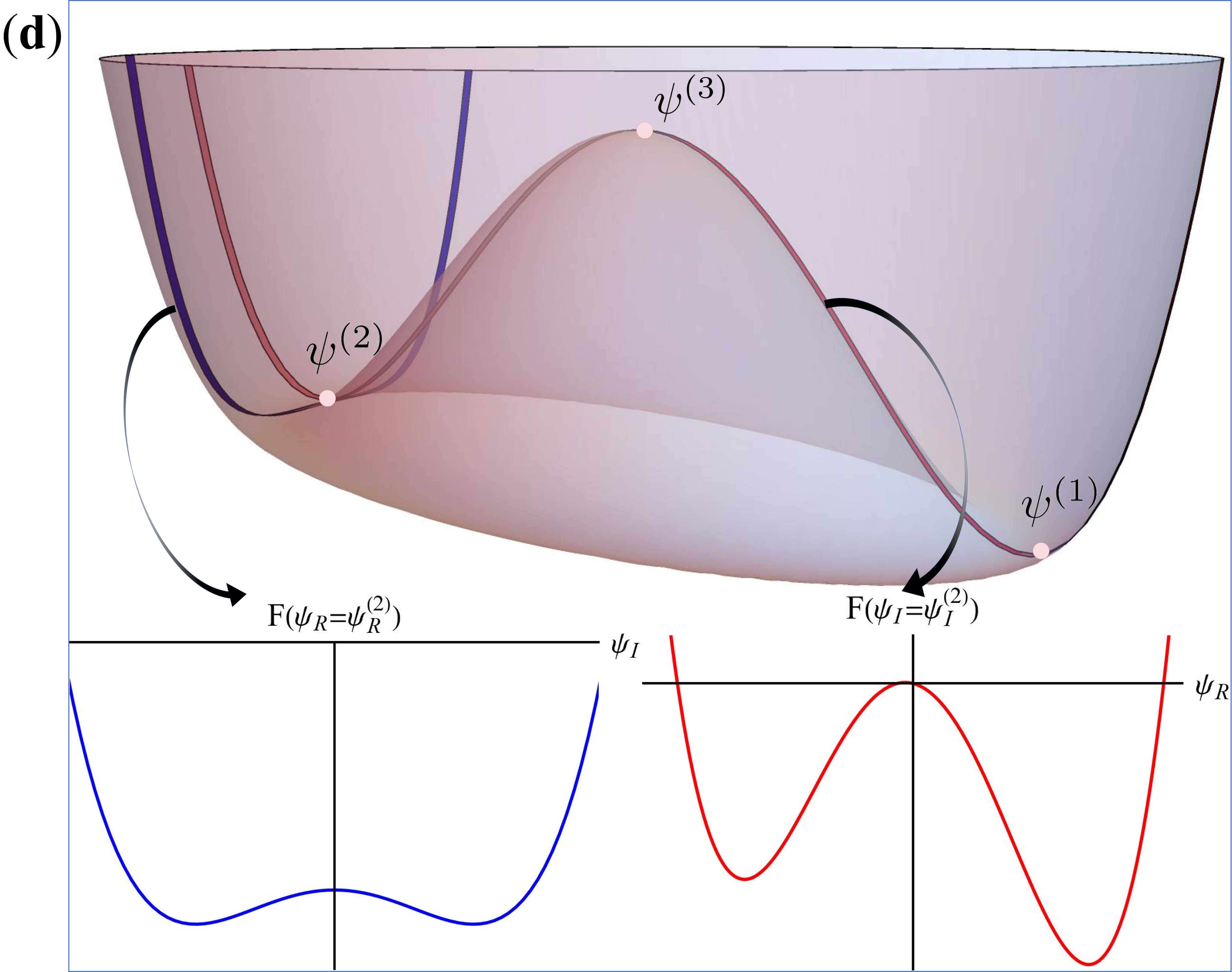}
\caption{The structure of the GL functional in Eq.~\eqref{pote} as a function of the real and imaginary parts of the order parameter $(\psi_R,\psi_I)$. In this figure, the external symmetry breaking field is fixed to a constant value $h=10^{-3}$ and $\beta=2$ for simplicity. The stationary points of the potential are indicated as $\psi^{(i)}$ with $i=1,2,3$. In panels (a)-(d) the parameter $\alpha$ is respectively fixed to $\alpha \approx [0.1,0,-0.0238,-0.1]$. The solid blue and red curves show respectively the cuts of the potential along the $\psi_R,\psi_I$ axes. Panel (c) corresponds to the critical value $\alpha_c$ below which three distinct stationary points appear. $\psi^{(1)}$ is always a minimum, $\psi^{(2)}$ is a saddle point and $\psi^{(3)}$ a maximum.}
\label{figpot}
\end{figure}
By investigating the stationary points of the potential in Eq.~\eqref{pote} in terms of the variables $\left(\psi_R,\psi_I\right)$, two distinct cases can be identified. For $\alpha$ larger than a certain critical (negative) value, that is analytically obtainable, the potential exhibits a single minimum with $\langle \psi \rangle \neq 0$. In the limit of a small external field $h \ll 1$ (with $\alpha \neq 0$), this minimum is located at $\left(\psi_R,\psi_I\right) \approx \left(\frac{h}{\alpha} + \mathcal{O}(h^2), 0\right)$ and corresponds to a potential value $F \approx -\frac{h^2}{\alpha}$. As $\alpha$ decreases, this minimum shifts further from the origin in the $\left(\psi_R,\psi_I\right)$ plane. This scenario is illustrated in panels (a) and (b) of Figure~\ref{figpot}.

Conversely, for $\alpha$ below the critical value $\alpha_c = -\left(\frac{27 \beta h^2}{4}\right)^{1/3}$ (see panel (c) of Figure~\ref{figpot}), a pair of new stationary points emerges at $\left(\psi_R,\psi_I\right) = \left(\sqrt{\frac{-\alpha}{\beta}} - \frac{h}{2\alpha} \pm \mathcal{O}(h^2), 0\right)$. Among these, the point with the smaller $\langle \psi \rangle$ is a local maximum, while the other is a saddle point, crucially, not a minimum (as in first-order phase transitions). This conclusion follows from the fact that the local curvature at this point is negative along the $\psi_I$ direction. This configuration is shown in panel (d) of Figure~\ref{figpot}.

\begin{figure}[htb]
\centering
\includegraphics[width=0.48\linewidth]{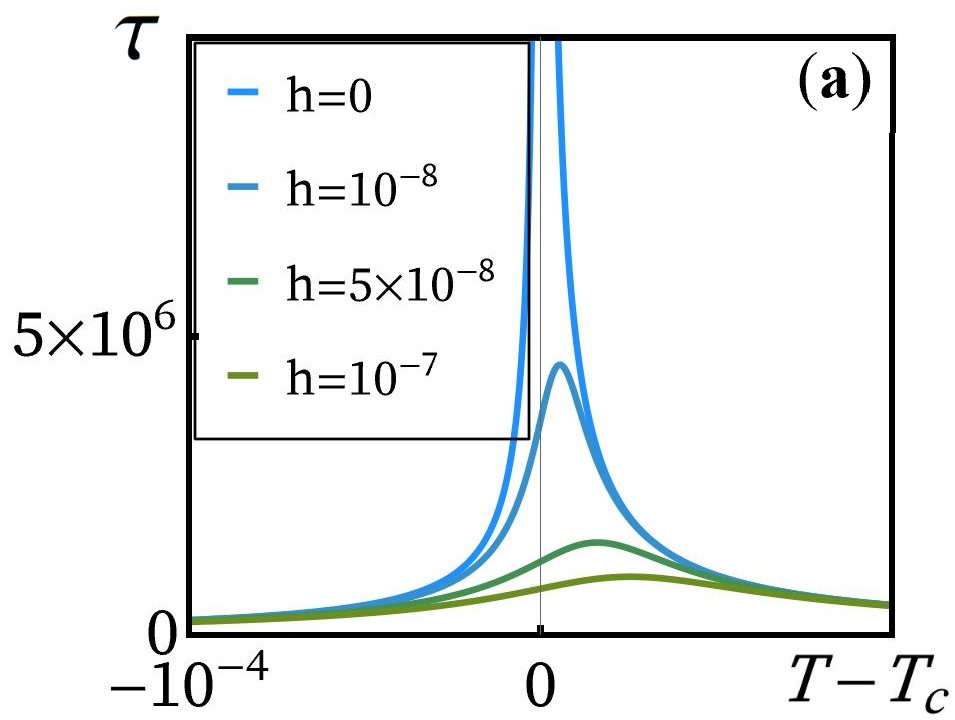}
\includegraphics[width=0.48\linewidth]{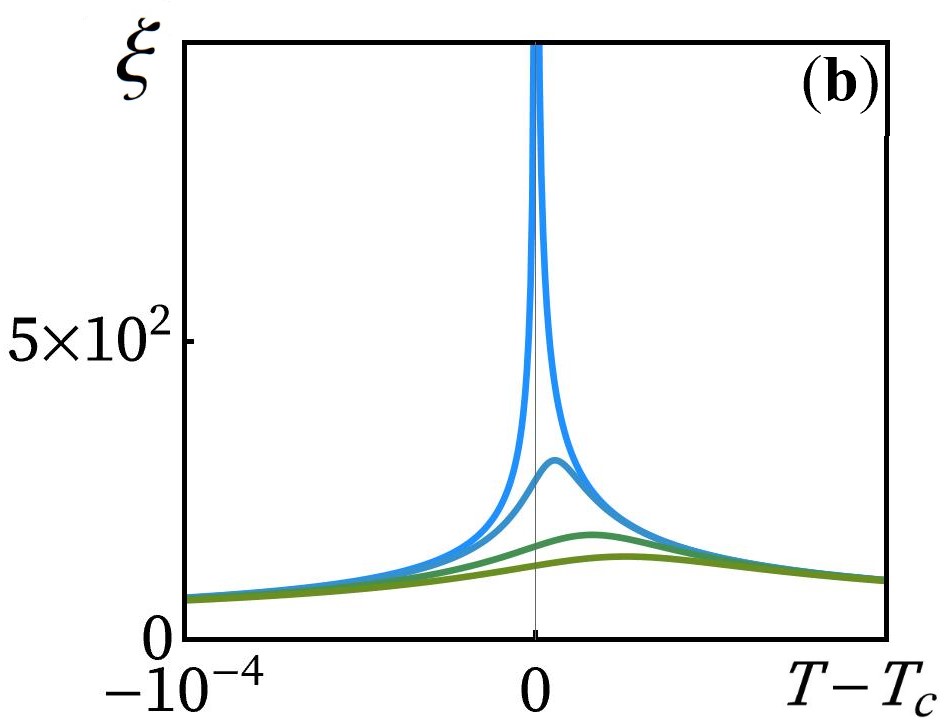}
\caption{Relaxation time $\tau$ \textbf{(a)} and correlation length $\xi$ \textbf{(b)} as functions of  $(T-T_c)\rightarrow0$ for different values of the symmetry-breaking parameter $h$.}
\label{cri}
\end{figure}

{\it Appendix B: Pseudo-critical dynamics -} The correlation length $\xi$ and relaxation time $\tau$ near a critical point can be extracted from the order-parameter correlation function~\cite{PhysRevD.98.106024}:
\begin{align}
    \langle \mathcal{O}(\omega,k)\mathcal{O}^\dagger(-\omega,-k) \rangle \sim \frac{1}{-i\tilde{c}\omega + k^2 + \frac{1}{\xi^2}},\\
    \xi(\omega=0) = \frac{1}{\Im[k^*]}, \qquad \tau(k=0) = \frac{1}{\Im[\omega^*]},
\end{align}
where $k^*$ and $\omega^*$ correspond to the lowest pole of the $\langle \mathcal{O} \mathcal{O}^\dagger \rangle$ correlator. In Fig.~\ref{cri}, we illustrate the pseudo-critical behavior of $\xi$ and $\tau$ as functions of the distance from the original critical point $T = T_c$, varied by tuning the symmetry-breaking parameter $h$. As expected, for finite $h$, the power-law divergence at $T = T_c$ is removed, and both $\tau$ (panel (a)) and $\xi$ (panel (b)) exhibit a smooth maximum that shifts away from $T_c$ as $h$ increases.

\begin{figure}[htb]
\centering
\includegraphics[width=0.48\linewidth]{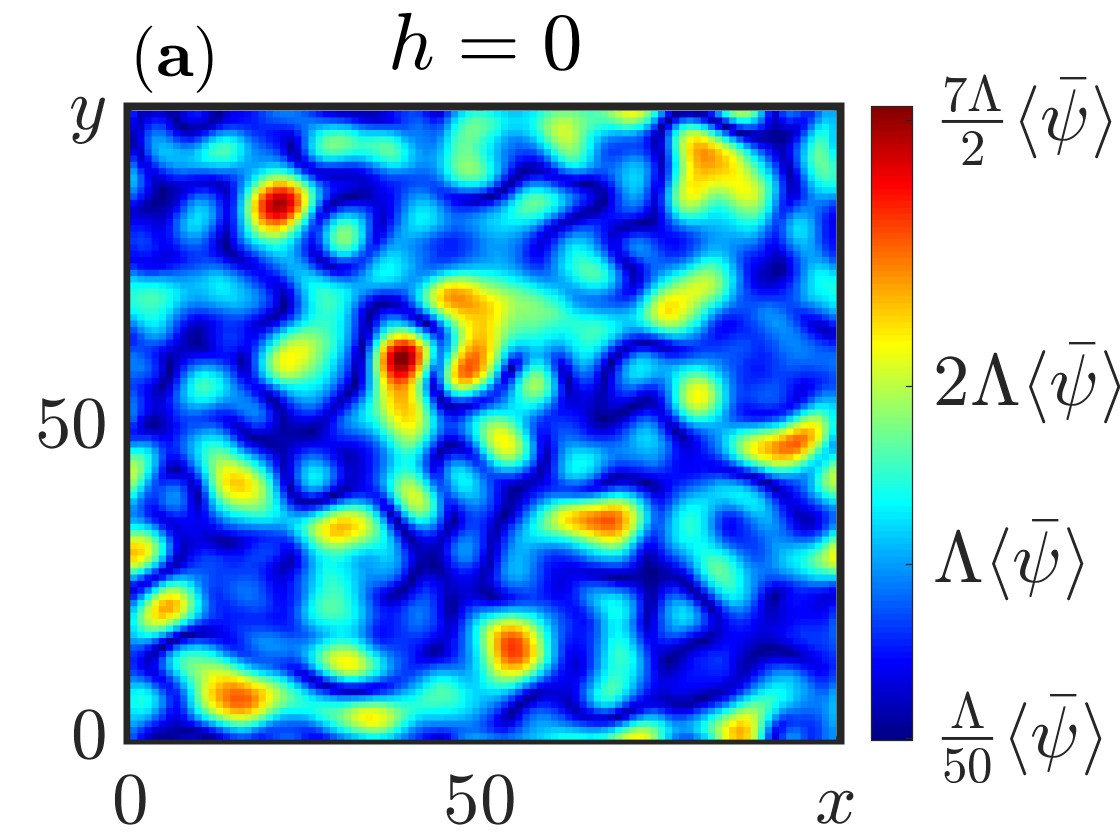}
\includegraphics[width=0.48\linewidth]{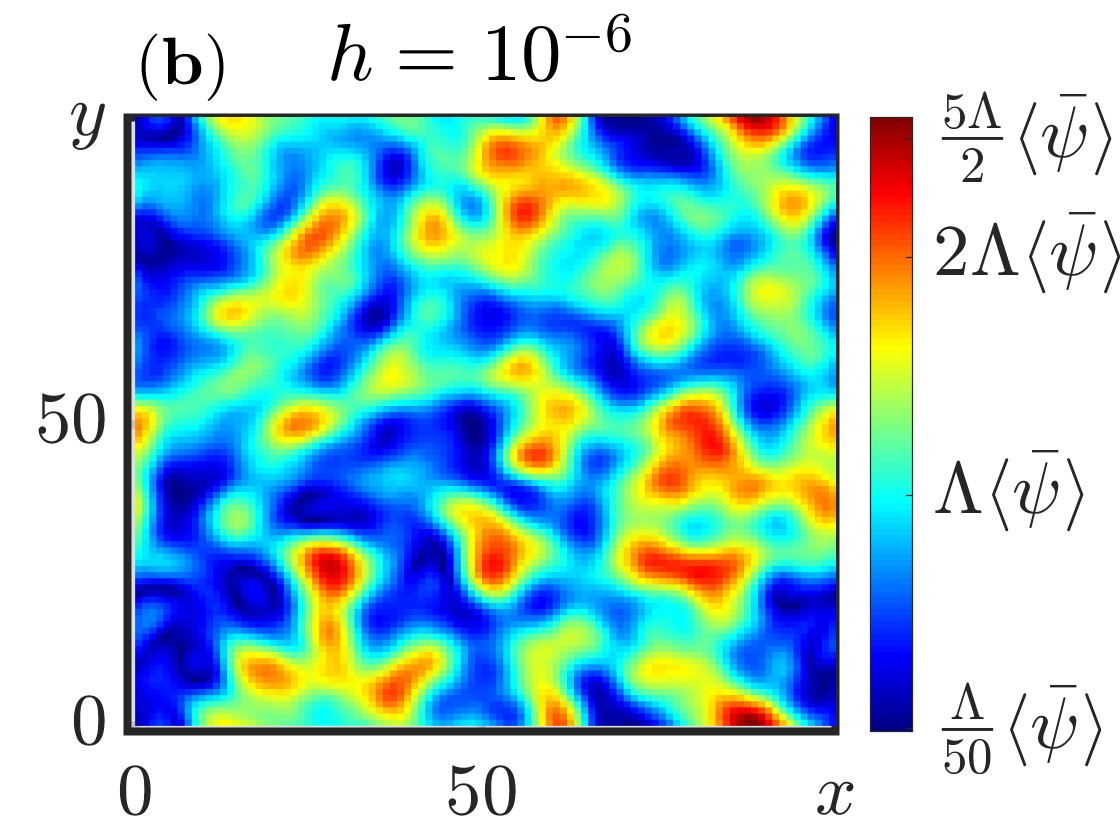}
\includegraphics[width=0.48\linewidth]{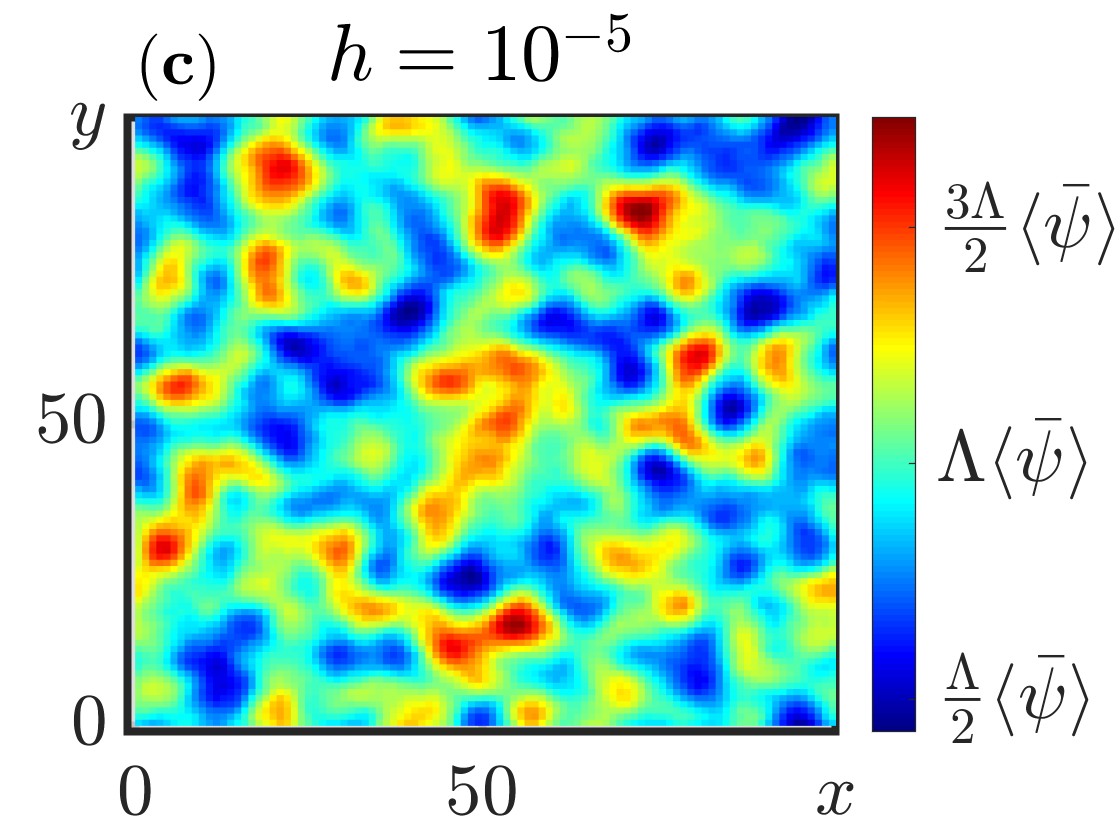}
\includegraphics[width=0.48\linewidth]{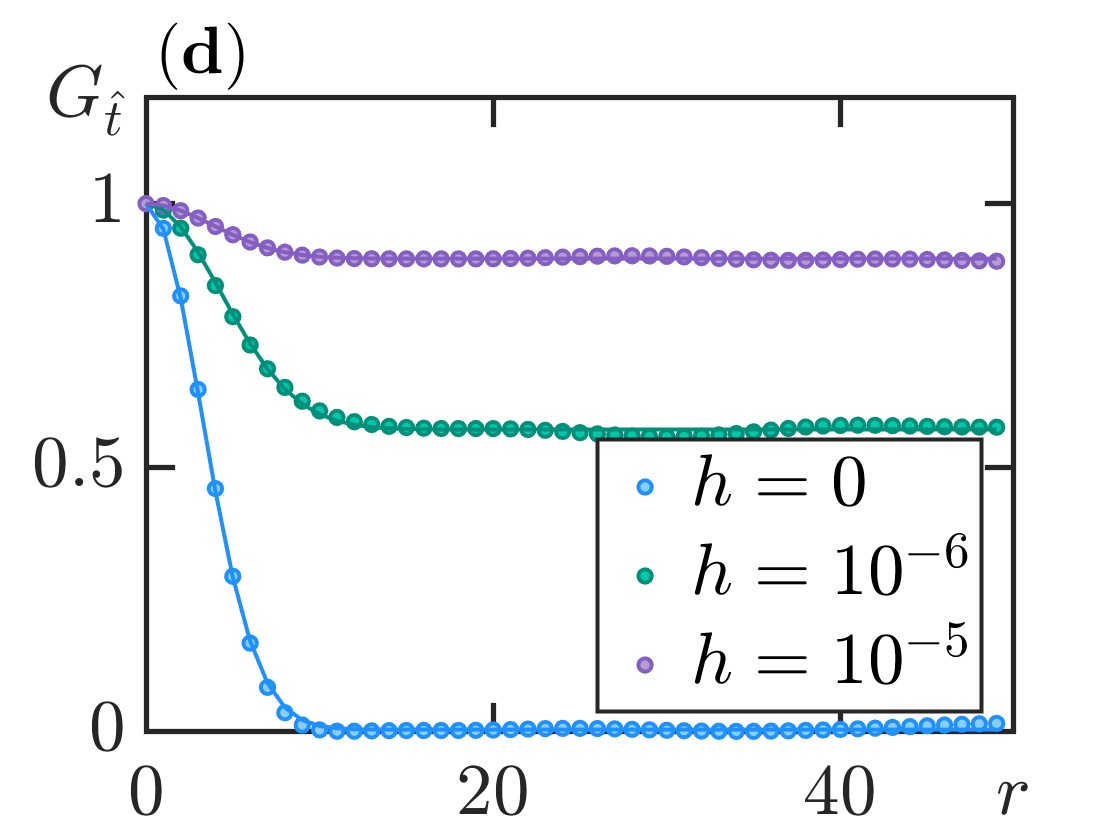}
\caption{\textbf{(a-c)} Spatial distribution of the order parameter $|\psi(x,y)|$ at freeze-out time with $h=0,10^{-6},10^{-5}$, respectively. \textbf{(d)} Radial correlation function for these three values of $h$. $\Lambda$ here is the cutoff used to define the freeze-out time (see SM).}
\label{fig_xi}
\end{figure}

{\it Appendix C: Spatial distribution of the order parameter -} Based on the definition of the density-density correlation function in Eq.~\eqref{eq.G}, we found in the main text that the correlation length increases with the parameter $h$. In Fig.~\ref{fig_xi}(a-c), we show three snapshots of the spatial distribution of the order parameter. It is visually evident that, as $h$ increases, the system forms larger clusters with uniform condensate, leading to an enhanced correlation length. This trend is further illustrated in Fig.~\ref{fig_xi}(d), where we plot the radial density-density correlation function. 

In the zero-source limit ($h = 0$), correlations decay rapidly and approach zero at large distances ($r \gtrapprox 10$). As $h$ increases, two distinct effects emerge: (i) the radial decay becomes progressively slower, and (ii) the correlation function no longer vanishes at infinity but instead saturates to a finite constant value that grows with $h$. These findings clearly confirm that increasing the symmetry-breaking parameter $h$ enhances the correlation length of the order parameter, in agreement with the observed reduction in the number of topological defects.

{\it Appendix D: Details of the holographic model} -- To explicitly break the $U(1)$ symmetry in a holographic superfluid \cite{Herzog:2009xv,PhysRevLett.101.031601}, a sourced scalar field is introduced, as investigated in \cite{Ammon:2021pyz}. The action for the matter fields is given by
\begin{equation}
    S_\mathcal{M} = \int d^4x \left[-\frac{1}{4}F_{\mu\nu}F^{\mu\nu} - |D\Psi|^2 - m^2|\Psi|^2 \right],
\end{equation}
where $D_\mu = \partial_\mu - iA_\mu$ is the covariant derivative, $\Psi$ is a complex scalar field with mass $m$ (hereafter set to $m^2 = -2$), and $A_\mu$ is the $U(1)$ gauge field with field strength tensor $F = dA$.

In the probe limit, the background spacetime is fixed to be an AdS$_4$ Schwarzschild black hole. In Eddington-Finkelstein coordinates, the metric reads
\begin{equation}
    ds^2 = \frac{1}{z^2}\left[-f(z)dt^2 - 2dtdz + dx^2 + dy^2 \right],
\end{equation}
where $f(z) = 1 - \frac{z^3}{z_h^3}$ and $z_h$ denotes the position of the black hole horizon. The AdS$_4$ conformal boundary is located at $z = 0$, and the horizon is defined by $f(z_h) = 0$. In what follows, we set $z_h = 1$, such that the Hawking temperature becomes $T = \frac{3}{4\pi}$.

The dynamics of the matter fields $\{\Psi, A_\mu\}$ are governed by the equations of motion:
\begin{align}
    &\nabla_\mu F^{\mu\nu} = J^\nu, \label{eq_A} \\
    &D^2 \Psi = -2\Psi. \label{eq_Psi}
\end{align}
Near the AdS boundary, the scalar field expands as
\begin{align}
    \Psi = \Psi^{(0)}z + \Psi^{(1)}z^2 + \mathcal{O}(z^3).
\end{align}

According to the holographic dictionary \cite{Zaanen_Liu_Sun_Schalm_2015,McGreevy2010,hartnoll2018holographic}, and using alternative quantization (consistent with previous literature \cite{Bu:2024oyz,Bu:2021clf}), $\Psi^{(0)}$ is interpreted as the order parameter $\langle O \rangle$ of the boundary superfluid condensate, with the $U(1)$ symmetry spontaneously broken. The boundary condition
\begin{equation}
    \partial_t \Psi^{(0)} - \Psi^{(1)} - i a_t \Psi^{(0)} = h
\end{equation}
introduces a nonzero source $h$ for $\langle O \rangle$, thereby explicitly breaking the $U(1)$ symmetry \cite{Ammon:2021pyz}.

\begin{figure}[htb]
\centering
\includegraphics[width=0.9\linewidth]{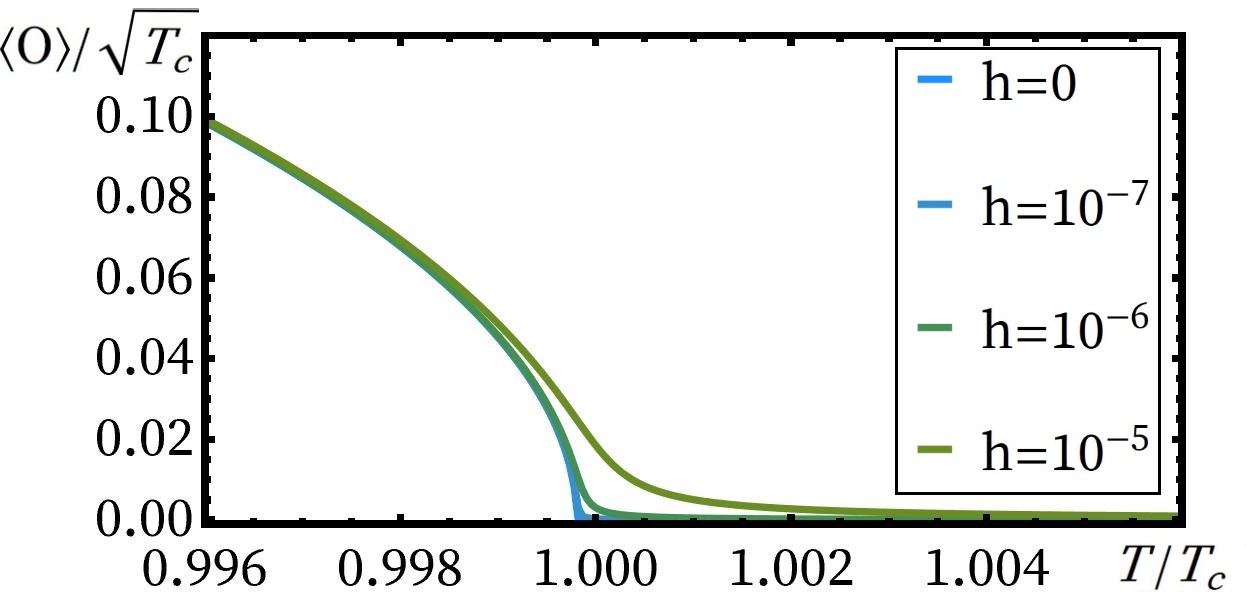}
\caption{Superfluid condensate as a function of temperature for different values of the symmetry-breaking source $h$.}
\label{holo}
\end{figure}

For completeness, in Fig.~\ref{holo}, we show the condensate as a function of temperature for various values of $h$.

To numerically simulate the dynamics of a holographic superfluid system undergoing a quench, one must solve the coupled equations of motion for the bulk fields $\Psi$ and $A_\mu$:
\begin{align}
\partial_t \partial_z \psi &= \partial_z\left(\frac{f(z)}{2}\partial_z\psi \right) + \frac{1}{2}\partial^2\psi - i\,\textbf{A} \cdot \partial\psi + iA_t \partial_z\psi \nonumber \\
&\quad - \frac{i}{2}\left(\partial\cdot\textbf{A} - \partial_z A_t\right)\psi - \frac{1}{2}\left(z + \textbf{A}^2\right)\psi, \label{dphi} \\
\partial_t \partial_z \textbf{A} &= \partial_z\left(\frac{f(z)}{2}\partial_z\textbf{A}\right) - |\psi|^2\textbf{A} + \textrm{Im}(\psi^* \partial\psi) \nonumber \\
&\quad + \frac{1}{2}\left[\partial\partial_z A_t + \partial^2 \textbf{A} - \partial\partial \cdot \textbf{A}\right], \label{eqeom} \\
\partial_t \partial_z A_t &= \partial^2 A_t - \partial_t \partial\cdot \textbf{A} + f(z) \partial_z \partial\cdot \textbf{A} - 2A_t |\psi|^2 \nonumber \\
&\quad + 2\,\textrm{Im}(\psi^* \partial_t \psi) - 2f(z)\,\textrm{Im}(\psi^* \partial_z \psi). \label{eqAt}
\end{align}

In the above equations, we define $\psi \equiv \Psi/z$ and $\textbf{A} \equiv (A_x, A_y)$. The bulk fields depend on time $t$ and spatial coordinates $(z, x, y)$, and we impose the following boundary conditions at the AdS boundary $z = 0$:
\begin{equation}
\partial_z A_t\big|_{z=0} = -\rho_c \left(1 - \frac{t}{t_Q} \right)^{-2}, \qquad \textbf{A}\big|_{z=0} = 0. \label{bcs}
\end{equation}

As in the Ginzburg-Landau case, the dynamics are integrated using a fourth-order Runge-Kutta method to model the Kibble-Zurek mechanism in the holographic setup.

\newpage
\onecolumngrid
\appendix 
\clearpage
\renewcommand\thefigure{S\arabic{figure}}    
\setcounter{figure}{0} 
\renewcommand{\theequation}{S\arabic{equation}}
\setcounter{equation}{0}
\renewcommand{\thesubsection}{SM\arabic{subsection}}
\begin{center}
    {\Large \bf Supplementary Material for ``Topological Defect Formation Beyond the Kibble-Zurek Mechanism in Crossover Transitions with Approximate Symmetries''}
\end{center}
\begin{center}
    {\large Peng Yang, Chuan-Yin Xia, Sebastian Grieninger, Hua-Bi Zeng, Matteo Baggioli}\\
    
    Corresponding Authors: \color{blue}zenghuabi@hainanu.edu.cn \color{black}, \color{blue} b.matteo@sjtu.edu.cn \color{black}
\end{center}
In this Supplementary Material, we provide further details about the numerical methods and the stability of our analysis.
\subsection*{Quenching protocol and numerical methods}

\begin{figure}[htb]
\centering
\includegraphics[width=0.65\linewidth]{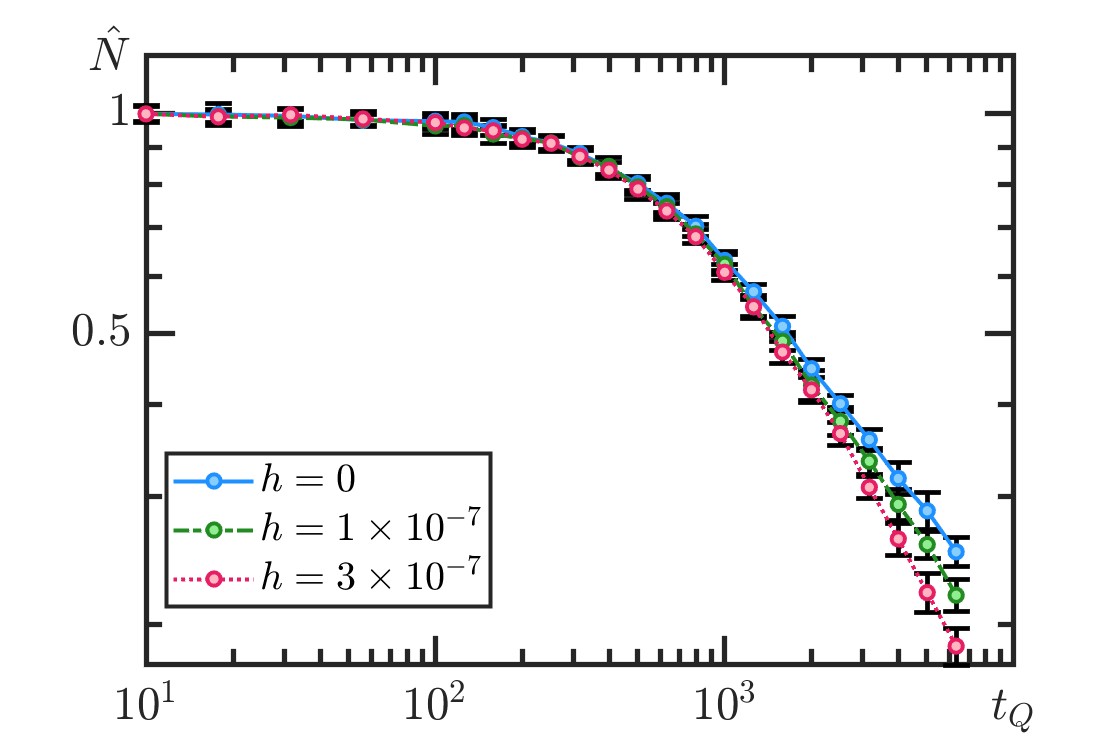}
\caption{\textbf{Numerical simulations for the number of topological defects}. Statistical average of the number of vortices over 50 independent simulations.}
\label{errorN}
\end{figure}

To simulate the Kibble-Zurek (KZ) mechanism in a two-dimensional system, we consider a periodic box of size $L = 500$. The time evolution from the initial state $\psi(t_0)$ to the final state $\psi(t_f)$ is computed by integrating the equation of motion given in the main text,
\[
\partial_t \psi = \mathcal{L}(\alpha(t), \psi(t)),
\]
using a fourth-order Runge-Kutta method:
\[
\psi(t) = \psi(t_0) + \int_{t_0}^{t} dt' \, \mathcal{L}(\alpha(t'), \psi(t')).
\]
The number of vortices is measured at the \textit{freeze-out} time $\hat{t}$, at which subdomains of the form $|\psi_i|e^{i\theta_i}$ are well-formed. The statistical average over $50$ independent simulations is shown in Fig.~\ref{errorN}.

\subsection*{Robustness of the results with respect to the definition of freeze-out time and initial configuration}
\begin{figure}[htb]
\centering
\includegraphics[width=0.48\linewidth]{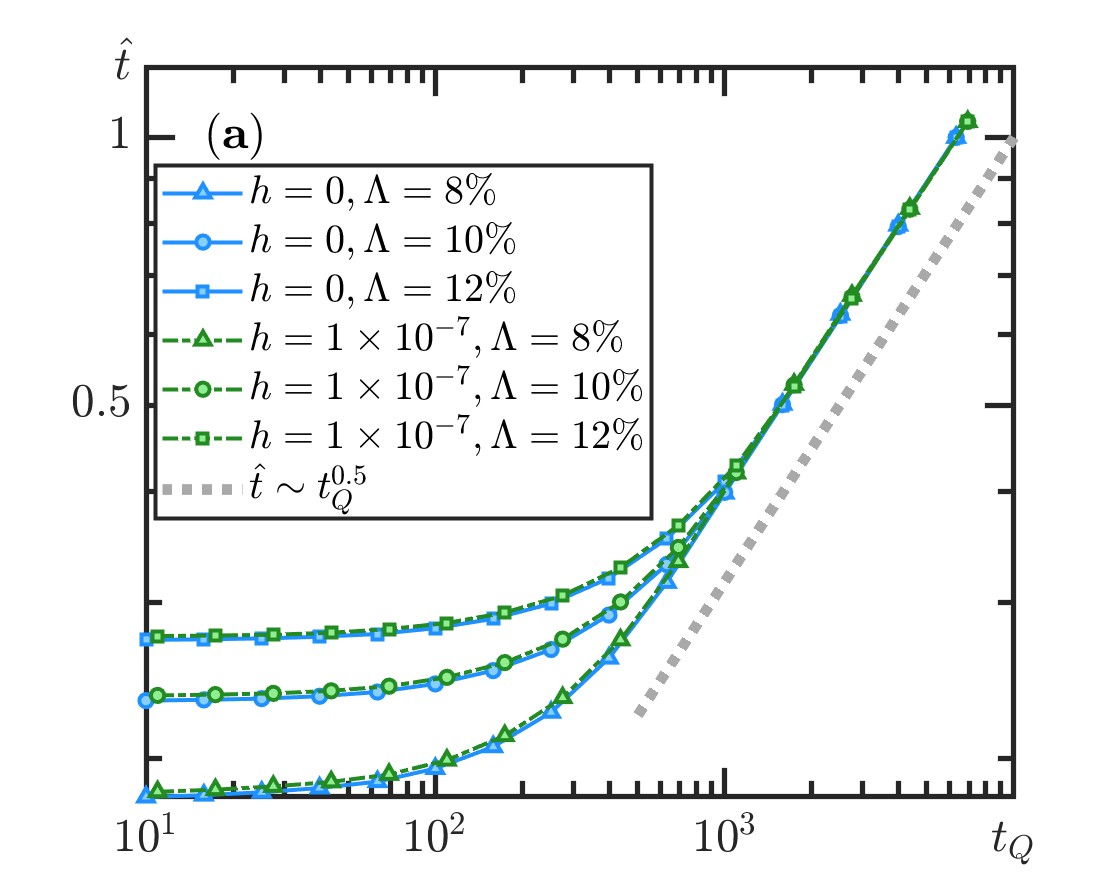}
\includegraphics[width=0.48\linewidth]{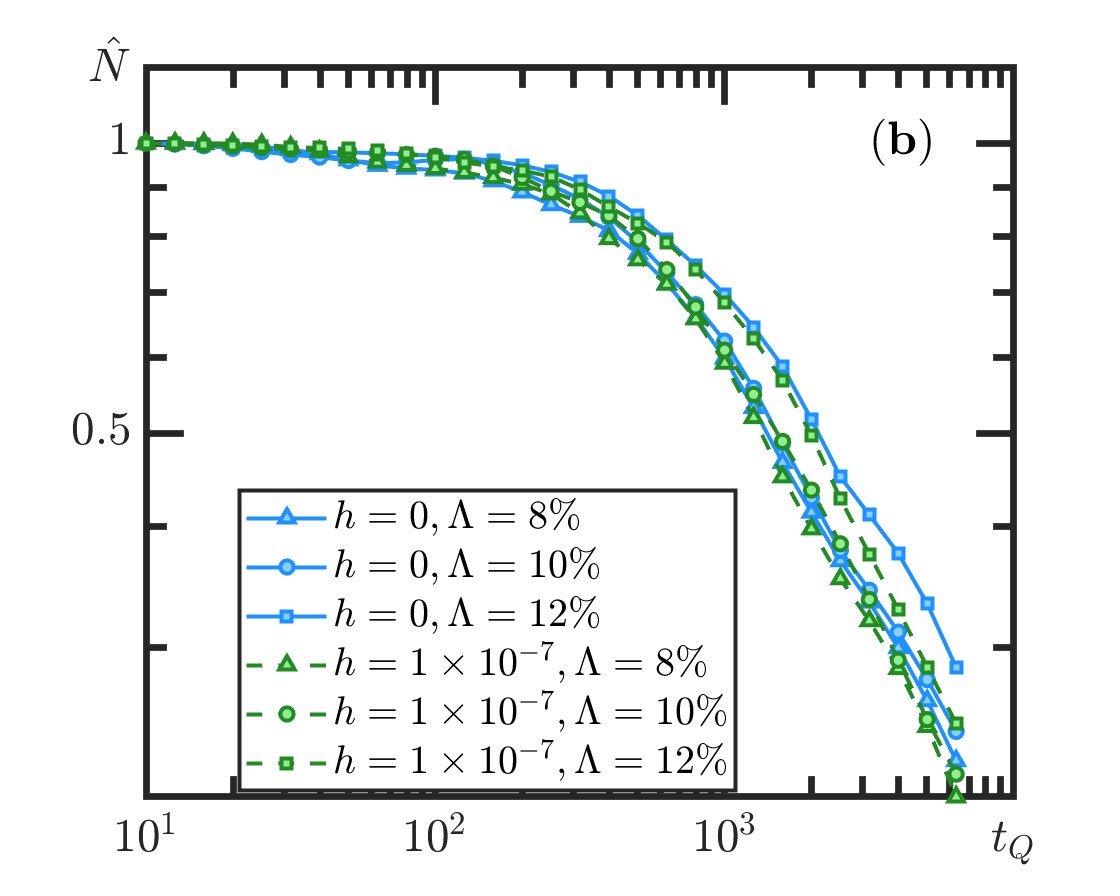}
\caption{\textbf{Testing the robustness of the results in terms of the freeze-out cutoff.} \textbf{(a)} $\hat{t}$ as a function of $t_Q$ for different choices of the numerical cutoff $\Lambda$. \textbf{(b)} Similar analysis for the normalized number of defects $\hat{N}$.}
\label{L}
\end{figure}

In the main text, we define the freeze-out time $\hat{t}$ by introducing a cutoff $\Lambda = 10\%$, corresponding to the time when the condensate $\langle \bar{\psi} \rangle$ reaches 10\% of its final value. In Fig.~\ref{L}, we demonstrate that the numerical results remain stable when varying the choice of $\Lambda \in[8\%,10\%,12\%]$.

From this analysis, we confirm that:
\begin{itemize}
    \item The freeze-out time $\hat{t}$ in the fast-quench regime is independent of the quench rate $t_Q$, as demonstrated in~\cite{PhysRevLett.130.060402};
    \item In the Kibble-Zurek (KZ) scaling regime, the freeze-out time $\hat{t}$ obeys the expected scaling behavior, i.e., $\hat{t} \sim t_Q^{0.5}$;
    \item Independently of the choice of the cutoff $\Lambda$, that defines the freeze-out time, the number of topological defects follows the modified KZ law, $\hat{N} \sim t_Q^{-1/2} e^{-\beta_h t_Q}$, as discussed in the main text.
\end{itemize}

\begin{figure}[htb]
\centering
\includegraphics[width=0.85\linewidth]{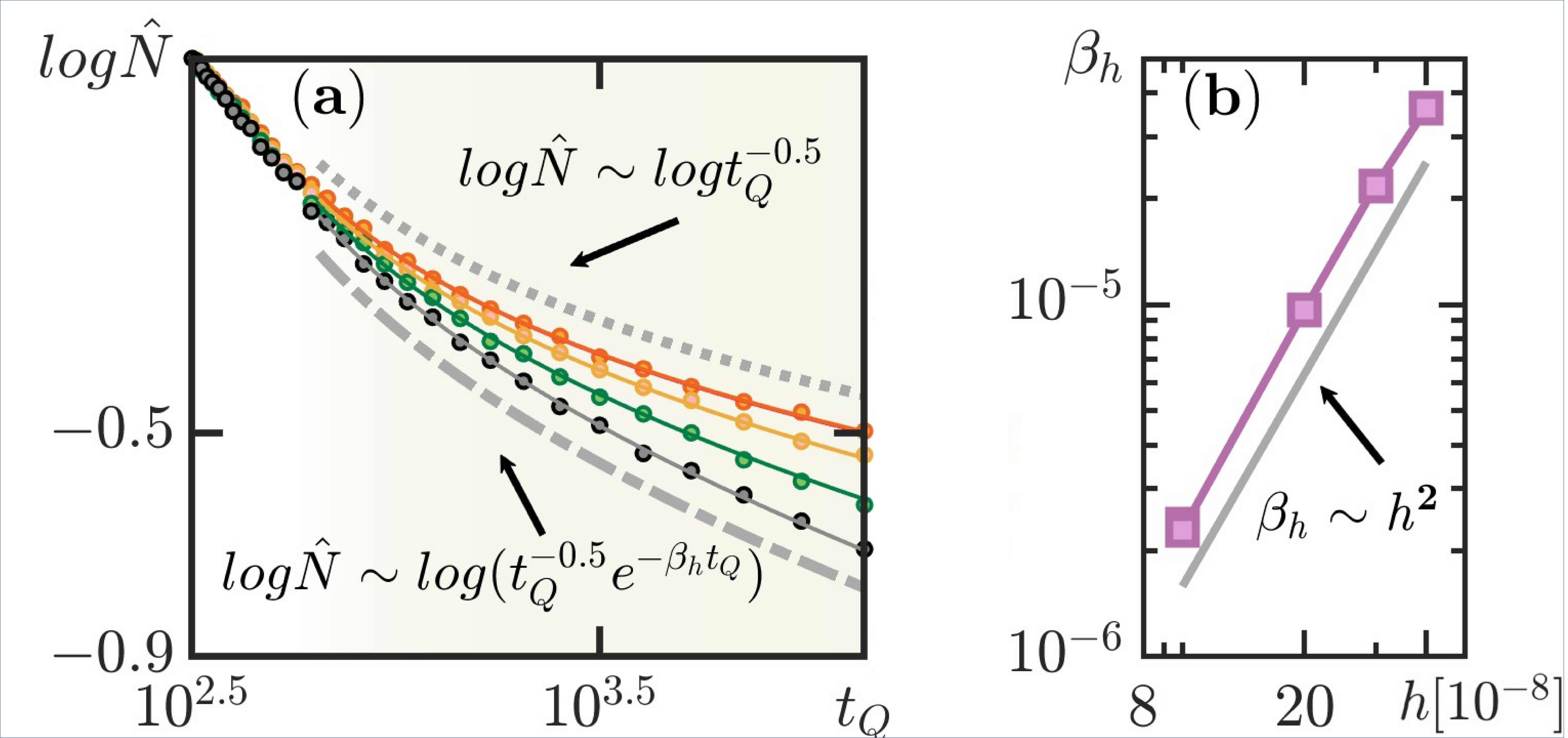}
\includegraphics[width=0.85\linewidth]{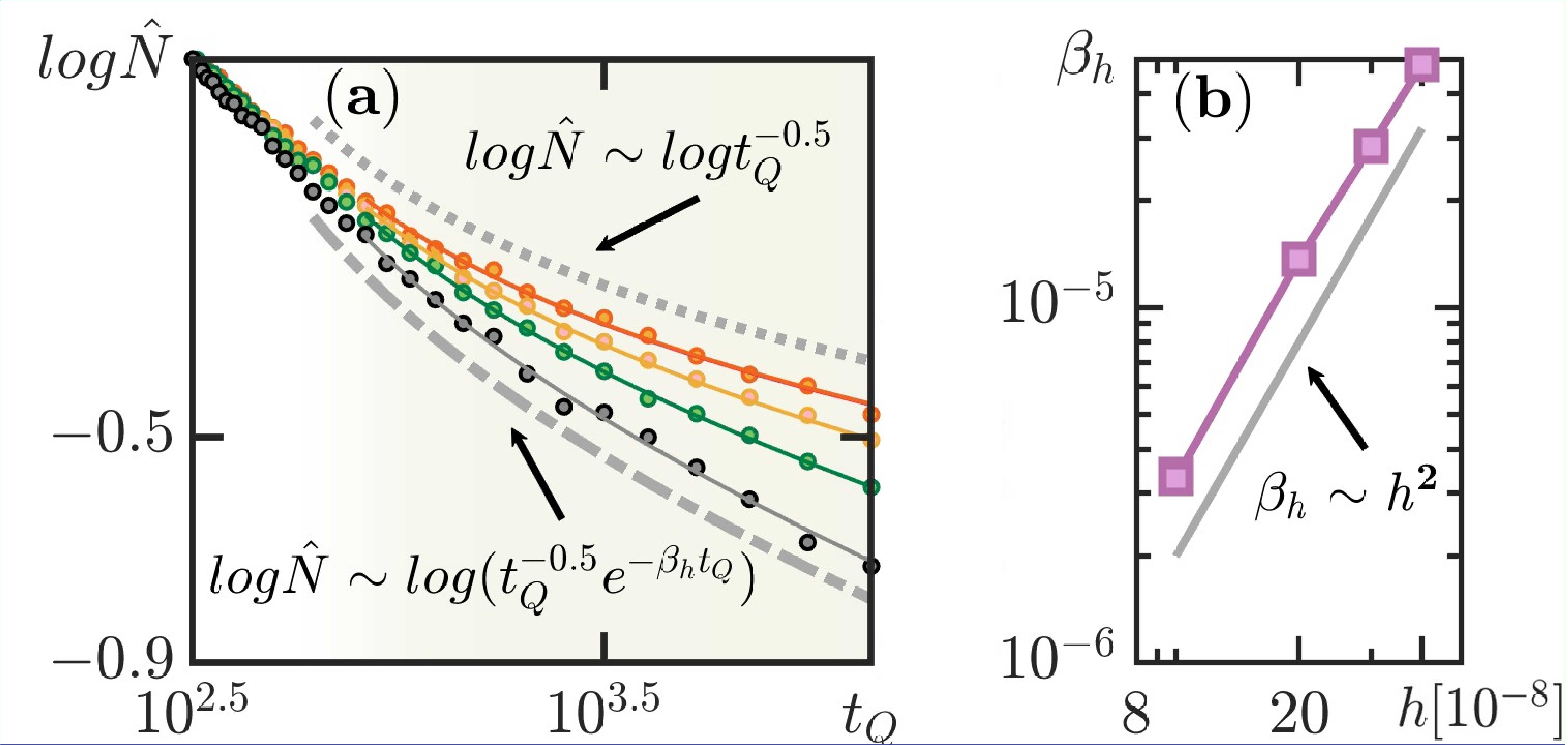}
\includegraphics[width=0.85\linewidth]{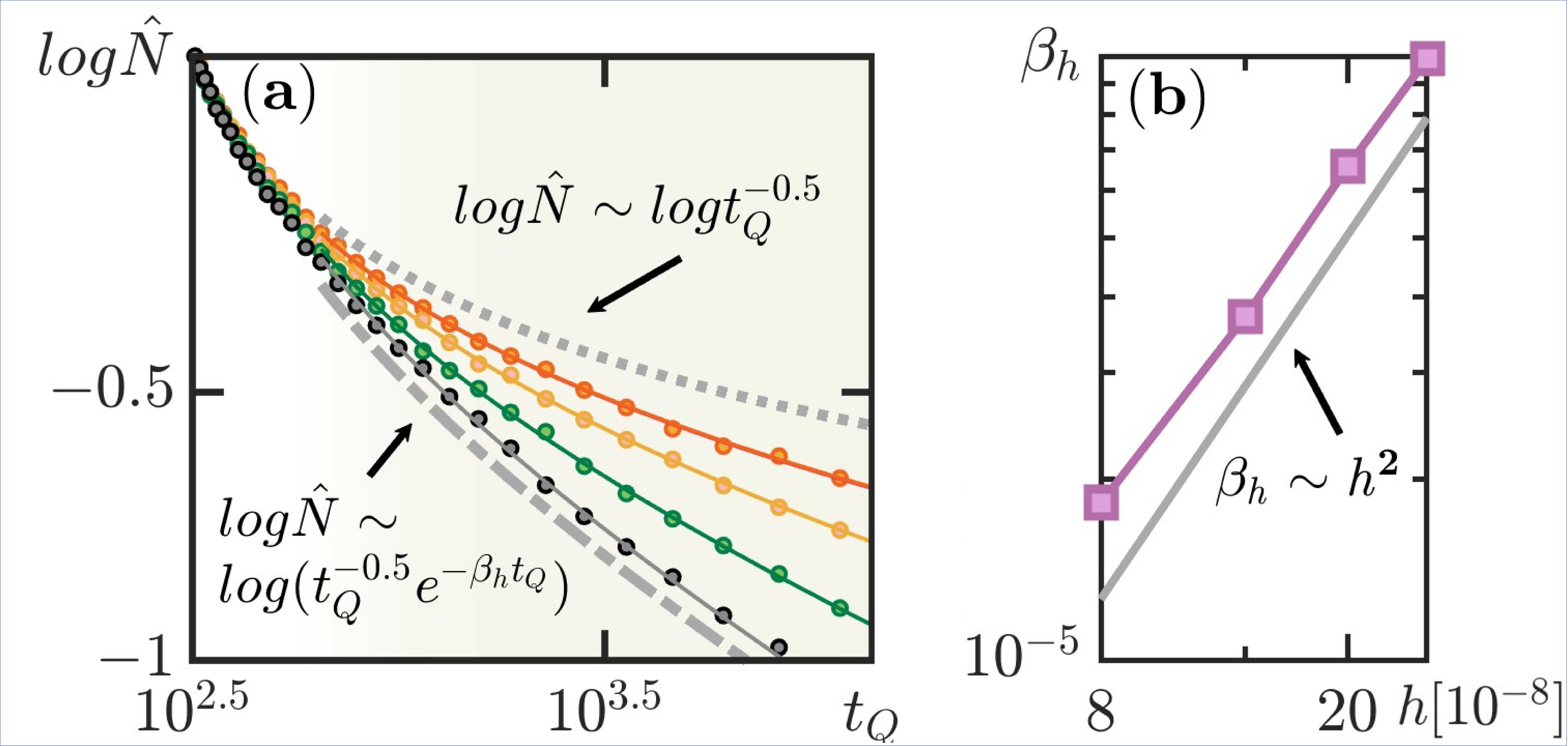}
\caption{\textbf{Robustness of the results upon changing the definition of the freeze-out time and the initial state. }\textbf{(a)} Zoom in the KZ scaling region for the normalized density of defects $\hat{N}$ as a function of the quench rate $t_Q$. Solid lines are fits of the numerical data using the generalized KZ form $\hat{N}\sim t_Q^{-0.5}e^{-\beta_h t_Q}$. \textbf{(b)} The behavior of $\beta_h$ as a function of the symmetry-breaking source $h$. \textbf{Top:} $\Lambda=10\%, \alpha(t=0)=0$.  \textbf{Middle:} $\Lambda=12\%, \alpha(t=0)=0$.  \textbf{Bottom:} $\Lambda=10\%, \alpha(t=0)=0.1$.}
\label{fig_S3}
\end{figure}

To examine this last point in greater detail, in Figure~\ref{fig_S3} we present an extended analysis of the number of defects as a function of the quench rate, varying both the freeze-out cutoff $\Lambda$ and the initial state from which the quench is performed. In all cases investigated, we find that the exponential correction is robust, and the coefficient $\beta_h$ consistently exhibits a quadratic scaling $\sim h^2$ with respect to the explicit symmetry-breaking external field $h$.

Overall, this extended analysis further reinforces the validity and robustness of our findings.

\end{document}